%% file: main.tex
\documentclass[nohyper]{JHEP3}
\usepackage{amsfonts,graphics,epsfig,amssymb}

\newcommand{\beq}{\begin{equation}}
\newcommand{\eeq}{\end{equation}}
\newcommand{\bea}{\begin{eqnarray}}
\newcommand{\eea}{\end{eqnarray}}
\newcommand{\beas}{\begin{eqnarray*}}
\newcommand{\eeas}{\end{eqnarray*}}

\def\Tr{ \hbox{\rm Tr}\,}

\def\SU{{\rm SU}}

\def\U{{\rm U}}

\def\tE{\widetilde{E}}

\def\Z{\mathbb{Z}}
\def\1{\mathbbm{1}}
\def\N{{\cal N}}

\def\W{{\cal W}}

\def\snd{ \mbox{\tiny $ \N=2$ } }
\def\snu{ \mbox{\tiny $ \N=1$ } }
\def\n{{\cal N}}
\def\0{\mbox{\tiny $(0)$ }}
\def\1{\mbox{\tiny $(1)$ }}
\def\2{\mbox{\tiny $(2)$ }}
\def\3{\mbox{\tiny $(3)$ }}
\def\e{{\rm e}}
\def\m{{\rm m}}
\def\W{{\rm W}}
\def\S{{\rm S}}

\title{ Strings Inside Walls in  $\N=1$ Super Yang-Mills}
\author{S. Bolognesi\\ 
William I. Fine Theoretical Physics Institute, University of Minnesota, 
116 Church St. S.E., Minneapolis, MN 55455, USA\\
\email{bolognesi@physics.umn.edu}}

\abstract{We conjecture the existence of strings bounded inside walls in
SU$(n)$ $\N=1$ Super Yang-Mills theory. These strings carry
$\Z_{[k,n]}$ quantum number, where $[k,n]$ is the greatest common
divisor between $k$, the charge of the wall, and $n$. We provide field-theoretical arguments and
string-theoretical evidences, both from MQCD and from gauge-gravity
correspondence. We interpret this result from the point of view of
the low-energy effective action living on the $k$-wall.
}


\preprint{FTPI-MINN-07/31; UMN-TH-2622/07}

\begin{document}

\section{Introduction}

In this paper we want to show that $\Z_{[k,n]}$ confining strings
live inside a $k$-wall of SU$(n)$ $\N=1$ SYM. From now on we shall
indicate as $[k,n]$ the maximum common divisor between $k$ and $n$.

We have a series of arguments to support our claim. We start in
Section \ref{heuristic} with an heuristic argument.  In
Sec.~\ref{MQCD} we give a more substantial argument from the MQCD
realization of the theory. In Sec.~\ref{gaugegravity} we discuss
the gauge-gravity setup, and how to understand our result. In Sec.~\ref{fieldtheory} we provide a field theoretical proof, using $\N=2$ SYM softly broken.
Here we will discuss the important analogy with the recent work \cite{Auzzi:2008zd}.  In
Sec.~\ref{objection} we discuss the peeling issue.  We then
move to the low energy effective theory in $2+1$ dimensions, in
Sec.~\ref{lowenergy}, and see why this effect can not be seen from the $2+1$ effective ation. At the end we summarize  our
results in Sec.~\ref{conclusion}. Appendix \ref{junction} contains a discussion about the string-wall junction.

%\newpage
\section{Heuristic Argument} \label{heuristic}
\input{heuristic}

%\newpage
\section{MQCD} \label{MQCD}
\input{MQCD}

%\newpage
\section{Gauge-Gravity Correspondence} \label{gaugegravity}

\input{gaugegravity}

%\newpage
\section{Field Theoretical Arguments} \label{fieldtheory}
\input{fieldtheory}

%\newpage
\section{Peeling} \label{objection}
\input{objection}

%\newpage
\section{Domain Wall Effective Action} \label{lowenergy}
\input{effectiveaction}

%\newpage
\section{Conclusion} \label{conclusion}

We now conclude, summarizing the main result of this paper. We
have argued that in the presence of a $k$-wall of SU$(n)$
super-Yang-Mills, there are  $\Z_{[k,n]}$ topologically stable
strings. According to the direction in which we orient the string
with respect to the wall we can have different scenarios. If the
strings are perpendicular, they cross the domain wall and continue
to the other half space. They can change their $n$-ality only modulo
$[k,n]$. If the strings are parallel to the wall, they will feel an
attractive force toward it. Their more energetically favorable
position will be inside the wall, as a bound state.

We have used various arguments to support our claim: one heuristic,
one from MQCD, another from the gauge-gravity correspondence and finaly a field theoretical one.

Still it is not completely clear what the nature of the low-energy effective action on the domain walls is.
Is not clear if the Acharya-Vafa theory is the right one or a more generic, maybe non-local, description is needed. 
The low-energy effective action would certainly be a usefull tool in the approach to these problems. But we should also keep in mind that the string inside wall phenomenon is not expected to be detectable in any $2+1$ effective action \cite{Auzzi:2008zd}. Although localized on the wall, it is still a fully $3+1$ effect.\footnote{The previous version of the paper contained an erroneous interpretation of the Acharya-Vafa theory in relation to the string inside wall phenomenon.}

The string-wall junction is not yet under quantitative control. The most crude approximation we can take has been described in Section \ref{fieldtheory}. Due to the non-locality of the fields, we are unable to make a global ansatz for the fields and study the junction in a solitonic, weakly coupled approach.

%\newpage
\appendix
\section{The String-Wall Junction}
\label{junction}

\input{junction}

%\newpage
\acknowledgments
I am grateful to R.~Auzzi, A.~Yung, and  M.~Shifman. for various discussions and the collaboration in Ref.~\cite{Auzzi:2008zd}.
I am very grateful to J.~Evlin for usefull discussion in Bruxelles in May 2007.
I want to thank D.~Tong for discussions at the Isaac Newton Institute
in September 2007. Part of this work has been done in
the Summer of 2007 while I was supported by the Marie Curie grant
MEXT-CT-2004-013510.  My work is now supported by DOE grant DE-FG02-94ER40823.

\end{document}

%% file: heuristic.tex
Let us start with some basic facts about $\N=1$ SU$(n)$ super
Yang-Mills. The theory has a U$(1)_R$ axial symmetry broken by an
anomaly to $\Z_{2n}$. This remnant symmetry is further broken to
$\Z_2$ by the gluino condensate $\langle \lambda \lambda \rangle
\propto n e^{i2\pi k/n}\Lambda^3$; the theory thus possesses
$n$ degenerate vacua labeled by a $\Z_n$ number. Two distinct vacua,
let us say $h$-vacuum and $(h+k)$-vacuum, can be separated by a
domain wall which we shall denote as a $k$-wall. These walls are
$1/2$ BPS saturated and their tension is equal to the modulus of the
difference of the superpotentials between the two vacua \cite{SD}.
Everyone of these vacua is in a massive phase where probe quarks are
confined. According to 't Hooft's classification of massive phases, we
must specify the $\Z_n^{\rm ele} \times \Z_n^{\rm mag}$ charges of the
particles that condense and are responsible for confinement. In the
$h$-vacuum of $\N=1$ SYM, confinement is due to the condensation of
an $(h,1)$ particle.

In $\N=1$ SYM there are thus two interesting extended objects, both
labeled by a $\Z_n$ number: domain walls and confining strings. We
shall now show that a bound state between these two objects exists.
A $\Z_{[k,n]}$ string can live inside a $k$-wall.

Something is already known about the relation between strings and
walls. For example, as first noted in \cite{WittenMQCD}, a
$1$-string can end on a $1$-wall. A simple field theoretical
argument goes as follows. A $1$-string can terminate on a
probe quark or on any object with charges $(1,0)$. A particle with
these charges can be created if the two condensates living on the
two sides of the wall can form a bound state. If, for simplicity, we
consider the $1$-wall separating the $0$ and the $1$ vacua, the two
condensates are, respectively, $(0,1)$ and $(1,1)$. A bound state
$-(0,1)+(1,1)$, an anti-monopole plus a dyon, would thus have the
charge of a fundamental quark and thus be a good ending point for a
$1$-string.

We want now to use the same argument and push it further, considering
a generic $h$-string perpendicular to a generic $k$-wall. Can this
string end, or not, on the domain wall? I would be possible if there
were a composite of the two condensates on which the $h$-string
could end. An $h$-string can end by definition on $h$ quarks, or on any
bound state with charge $(h,0)$. To be general, we consider the
$k$-wall that interpolates between the $p$ and the $p+k$ vacua, where
the two condensates responsible for confinement are, respectively,
$(p,1)$ and  $(p+k,1)$. We thus want to solve the following equation
for $a$ and $b$: \beq \label{charge} (h,0) = a (p,1) + b (p+k,1) \eeq
The equation is defined in the ring $\Z_n^{\rm ele} \times \Z_n^{\rm mag}$.
If the equation is solvable, the string can end on the wall; if not,
the string can not end on the wall and is forced to continue further
(see Figure \ref{primo}). From the monopole charges, the second in
$(.,.)$, we get $a=-b$, and from the electric charges we get: \beq  h=bk
\quad \mathrm{modulo} \quad n \eeq A simple theorem from  arithmetic
tells us that this equation is solvable if, and only if $h$ is a multiple
of the greatest common divisor between $k$ and $n$.
\FIGURE[h]{
\includegraphics[width=32em]{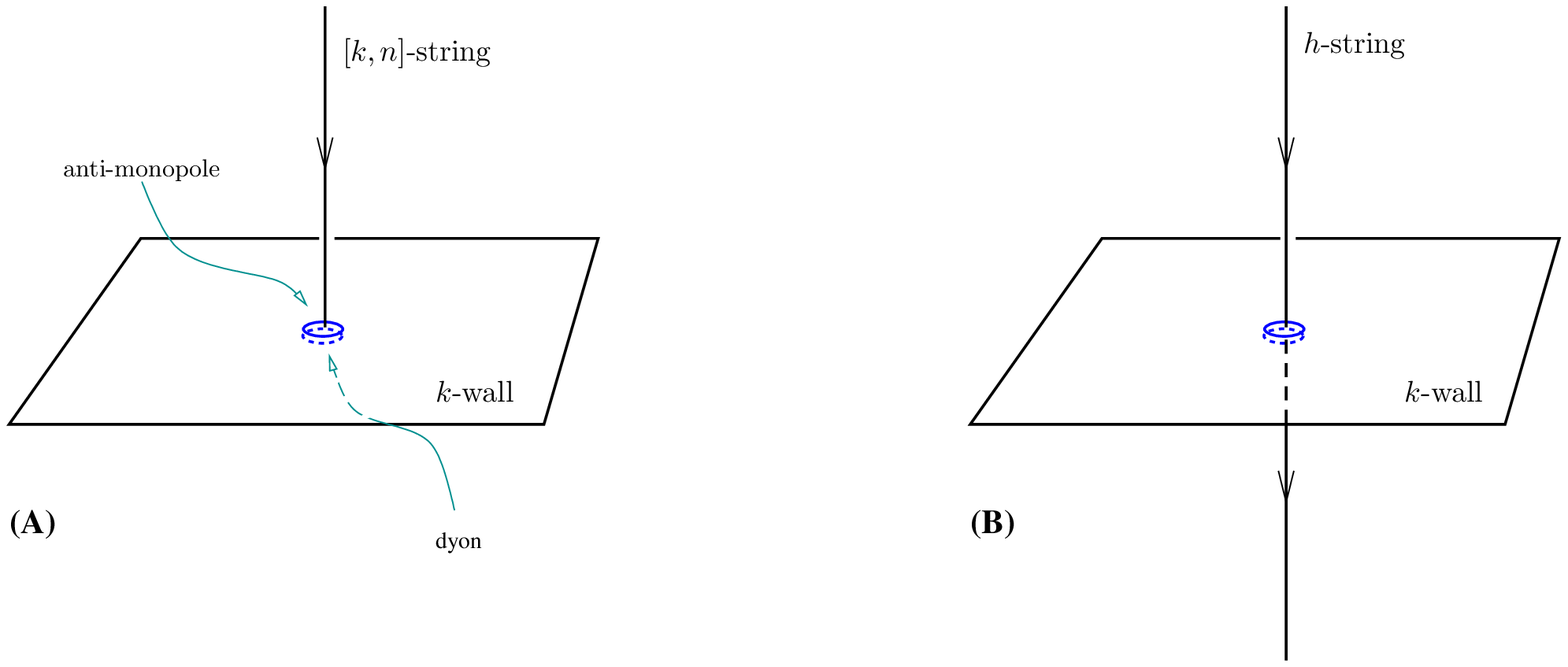}
\caption{ (A): An $h$-string can end on a
$k$-wall only if $h$ is a multiple of $[k,n]$. (B): If $h$ is not
divisible by $[k,n]$, the $h$-string can not terminate on the wall
and is forced to continue on the opposite vacuum.}
\label{primo}
}

We have thus the following scenario. When a $k$-wall interpolates
between a $p$ and a $p+k$ vacua, the group of $\Z_n$ strings is
divided into two categories. The subgroup $\Z_{n/[k,n]}$ of
strings multiple of $[k,n]$, can terminate on the wall. The quotient
group $\Z_{[k,n]}$ is instead topologically stable. Another way to
say this is that a string crossing the $k$-wall can change its $n$-ality
but can not change its $[k,n]$-ality.

The natural question is now what happens if the string, instead of
being perpendicular to the wall, is {\it parallel} to the wall. A
string in general feels an attractive force toward a parallel domain
wall. For example, a $1$-string is attracted toward a $1$-wall and
then dissolved. We thus conclude that any $h$-string parallel to a
$k$-wall feels an attractive force. When the $h$-string arrives on
the domain wall its fate is determined by the divisibility of $h$.
If $h$ is a multiple of $[k,n]$, the string is dissolved inside the
wall and disappears. If not, the string survives as a $\Z_{[k,n]}$
string bounded inside  the domain wall.

The other two deformations, described in Figure \ref{secondo}, reveals
the existence of this bound state of confining strings inside the
domain wall.
\FIGURE[h]{
\includegraphics[width=32em]{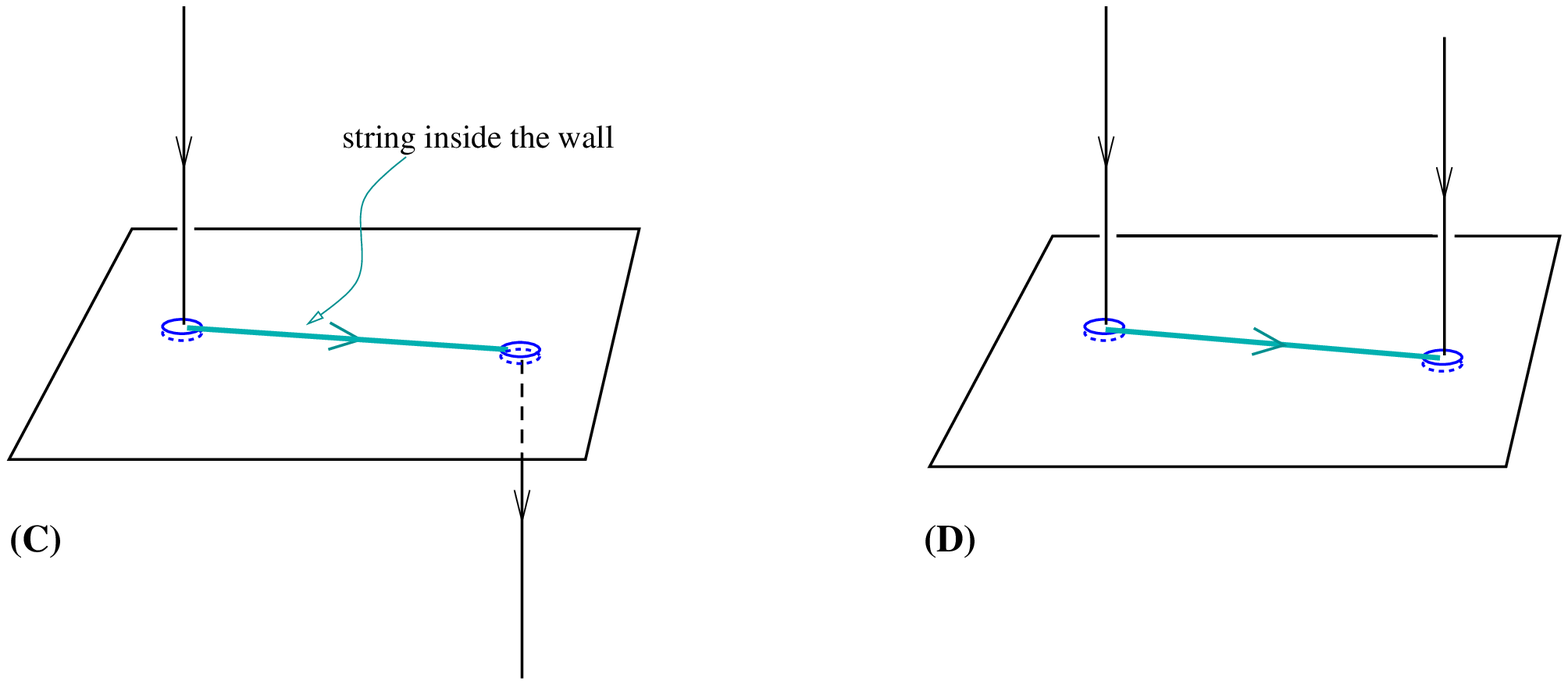}
\caption{ $(C)$: Figure $(B)$ of \ref{primo}
is modified so that the strings in the two vacua end on different
points on the domain wall. For consistency there should be a string,
living inside the domain wall, that connects the two ending points.
$(D)$: An $h$-string and a $-h$-string and on the $k$-wall on two
different points. If $h$ is not a multiple of $[k,n]$, a string
inside the domain wall must be created in order to connect the
ending points.}
\label{secondo}
}

%% file: MQCD.tex
Let us consider now the MQCD realization of $\N=1$ SYM. For a
detailed description we refer to the original papers
\cite{WittenMQCD,HSZ}; in the following we just present what is
needed to understand our result in the MQCD context.

We have M-theory compactified on a $S^1$ circle. The coordinates are
$x^0, \dots ,x^{9}$ and $x^{10}$, the M-theory circle, of period
$2\pi$. We define the complex coordinates $v=x^4+ix^5$, $w=x^7+ix^8$
and $t=e^{-(x^6 + ix^{10})}$. The gauge theory of interest is the
low energy limit of the effective action living on the world volume
of an M$5$-brane. The $5$-brane is extended on the four dimensional
space-time $x^{0},\dots,x^{3}$ times a non-compact Riemann surface
$\Sigma_p$. The Riemann surface is defined by the following
equations: \beq v^n = t \ , \qquad w=\zeta_p v^{-1} \ . \eeq where
$\zeta_p$ is the root of unity $e^{i2\pi p/n}$. The Riemann
surface, $\Sigma_p$, encodes certain information about the gauge
theory. Roughly, it can be seen as two infinite planes, the $v$ plane at
$x^{6}\to -\infty$ and the $w$ plane at $x^6 \to +\infty$, connected
by a tube that winds $n$ times around the circle of M-theory. There
are $n$ ways to connect the two planes depending on the choice of
the root of unity $\zeta_p$. These correspond to the $n$ discrete
vacua of the gauge theory (see Figure \ref{sigma} for an example).
\FIGURE[h]{
\includegraphics[width=34em]{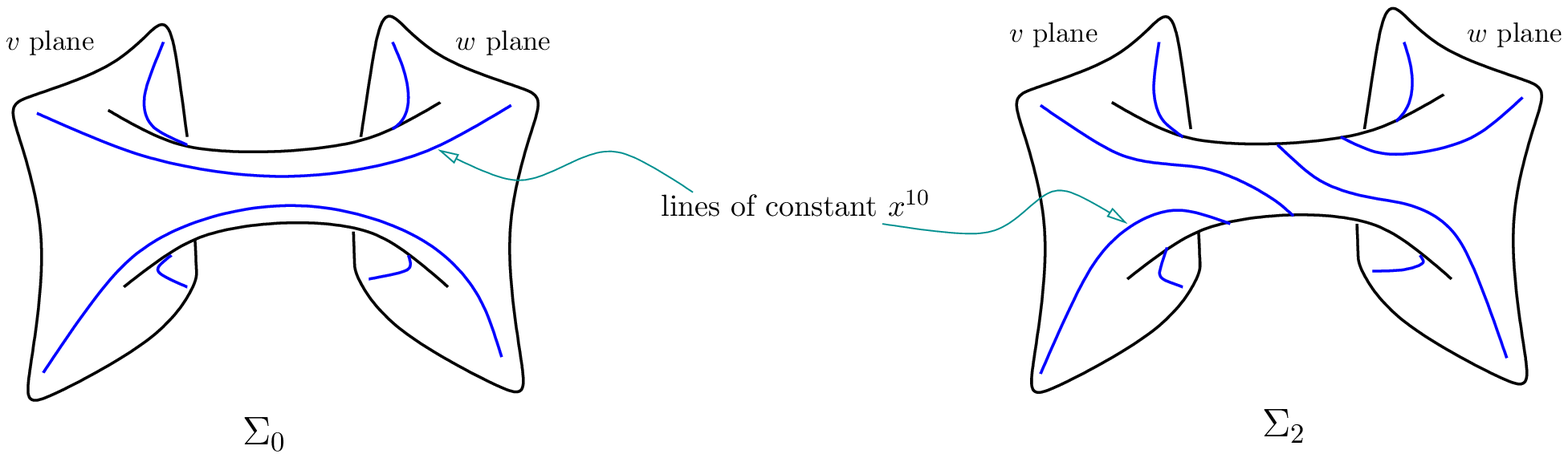}
\caption{ We provide two examples of Riemann
surface for $n=4$ (we choose this since the $2$-wall in SU$(4)$
$\N=1$ is the first non-trivial case in which our phenomenon
happens). In the figure we have $\Sigma_0$ and $\Sigma_2$
corresponding respectively to the vacua $0$ and $2$. The lines
represent the phase in the M-theory circle. The domain wall between
these two vacua is represented in Figure \ref{wall}. Note that the
$v$ and $w$ plane, despite what it seems from the figure, are not parallel, but orthogonal.}
\label{sigma}
}

Another feature of the gauge theory, nicely visible in this M-theoretical framework, is the presence of confining $\Z_n$ strings.
They correspond to M$2$-branes with $1+1$ dimensions extended in
space time, and the other spatial dimension extended on a finite
segment whose end points lie on the surface $\Sigma_p$. We call, for
convenience, $Y$ the $6$ dimensional manifold $R^{5}\times S^1 =
x^{4,\dots,8},x^{10}$. The Riemann surface, $\Sigma_p$, can thus be
thought of as embedded in this $6$ dimensional space $\Sigma_p \subset
Y$. Finite segments in $Y$, whose end points are forced to be on
$\Sigma_p$, correspond to elements of the relative homology group
$H_1(Y/\Sigma_p,\Z)$. From the exact sequence, we know how to express
this relative homology group as a function of the homology groups of
the $Y$ manifold and of the $\Sigma_p$ surface: \beq
H_1(Y/\Sigma_p,\Z)\cong \frac{H_1(Y,\Z)}{i(H_1(\Sigma_p,\Z))} \ .\eeq
$H_1(Y,\Z)$ is equal to $\Z$  simply counts the windings around the M
theory circle. $H_1(\Sigma_p,\Z)$ is also equal to $\Z$, and the unit
element consists of a winding around the tube connecting the $v$
plane with the $w$ plane. After the immersion $i(\;)$ into $Y$, this
unit element corresponds to $n$ windings around the M-theory circle.
We thus conclude that: \beq H_1(Y/\Sigma_p,\Z) = \\Z_n \ .\eeq

The fate of strings in the presence of a domain wall is again
determined by a relative homology group. But now we have to extend
the spaces in order to take into account of the $x^3$ direction
along which the wall interpolates between the two vacua. We define
the space $\widetilde{Y}$ as the $7$ dimensional manifold $x^3
\times Y$. The M$5$-brane extends on a three dimensional subspace
$S_k \in \widetilde{Y}$. $S_k$ has the following features. At $x^3
\to -\infty$ it approaches the surface $\Sigma_p$ while at $x^3 \to
+\infty$ it approaches the surface $\Sigma_{p+k}$. $S_k$ is thus a domain wall
that interpolates between the $p$ vacuum and the $p+k$ vacuum.
Strings in the domain wall background are represented by elements of
the relative homology group $H_1(\widetilde{Y}/S_k, \Z)$. Again, from
the exact sequence we can derive the formula: \beq
H_1(\widetilde{Y}/S_k,\Z)\cong
\frac{H_1(\widetilde{Y},\Z)}{i(H_1(S_k,\Z))} \ . \eeq
$H_1(\widetilde{Y},\Z)$ is equal to $\Z$ and is still the winding
around the M-theory circle. The task is now to evaluate $H_1(S_k,\Z)$
and embed it into the space $\widetilde{Y}$. We start with the
$1$-wall already considered in \cite{WittenMQCD}. $S_1$ has two
non-trivial cycles, $H_1(S_1,\Z)=\Z^2$. The first one is the same as
the Riemann surfaces $\Sigma$s. It is a circle that winds around the
tube connecting the $v$ plane and the $w$ plane. After embedding, it
corresponds to $n$ times the unit cycle of $H_1(\widetilde{Y},\Z)$.
The other non-trivial cycle is peculiar to the domain wall and is
constructed as follows. We divide the cycle into five distinct open
segments and then we connect them to form a closed circle. The first
four pieces make a very large loop in the $x^3,x^6$ plane, constant
in the M-theory circle $x^10$ and not closed in the $v$ plane. The
fifth and last piece closes the curve and winds once around $x^{10}$.
All the following movements are done staying inside the manifold
$S$. The construction, which can be seen in Figure \ref{wall}, goes
as follows. We start at $x^{3}\to -\infty$ in the $\Sigma_p$ vacuum
and at $x^{6} \to +\infty$ in the $w$ plane. We then move toward
$x^{6} \to -\infty$ into the $v$ plane on a line with a fixed $x^{10}$
coordinate. This is the first movement. The second step is to move
toward $x^{3} + \infty$, in the $\Sigma_{p+1}$ region, keeping fixed
$x^{6}$, $v$ and the $x^{10}$ phase. The third stage is to move back
into the $x^{6} \to +\infty$ region keeping $x^3$ fixed and the
phase $x^{10}$ constant. In the fourth passage we return to the
vacuum $\Sigma_p$ closing the circle in the $x^3, x^6$ plane. The
coordinate $w$ has been rotated by a phase $e^{\frac{i2\pi}{n}}$. To
close the circle we make the fifth and last movement, keeping $x^3$
and $x^6$ both fixed and rotating around the $w$ plane. This
inevitably makes a winding in the $x^{10}$ circle. These five steps
connected together form a closed circle, lying entirely in the
manifold $S_1$, that rotates once in the M-theory circle. This is
enough to conclude that $H_1(\widetilde{Y}/S_1,\Z)=1$, which implies
that every string can be ``unwinded'' in the domain wall background.
The same argument can be repeated for a generic $k$-wall $S_k$
interpolating between the vacua $\Sigma_p$ and $\Sigma_{p+k}$. $S_k$
again has  two non-trivial cycles. The first one is the same as the
Riemann surfaces $\Sigma$s and rotates $n$ times around the M
circle. The second is constructed in the same way as before with the
only difference being that it rotates $k$ times around $x^{10}$ instead of
only once (see Figure \ref{wall}). The relative homology group in
the $k$-wall background is thus: \beq
H_1(\widetilde{Y}/S_k,\Z)=\Z_{[k,n]} \ . \eeq
\FIGURE[h]{
\includegraphics[width=24em]{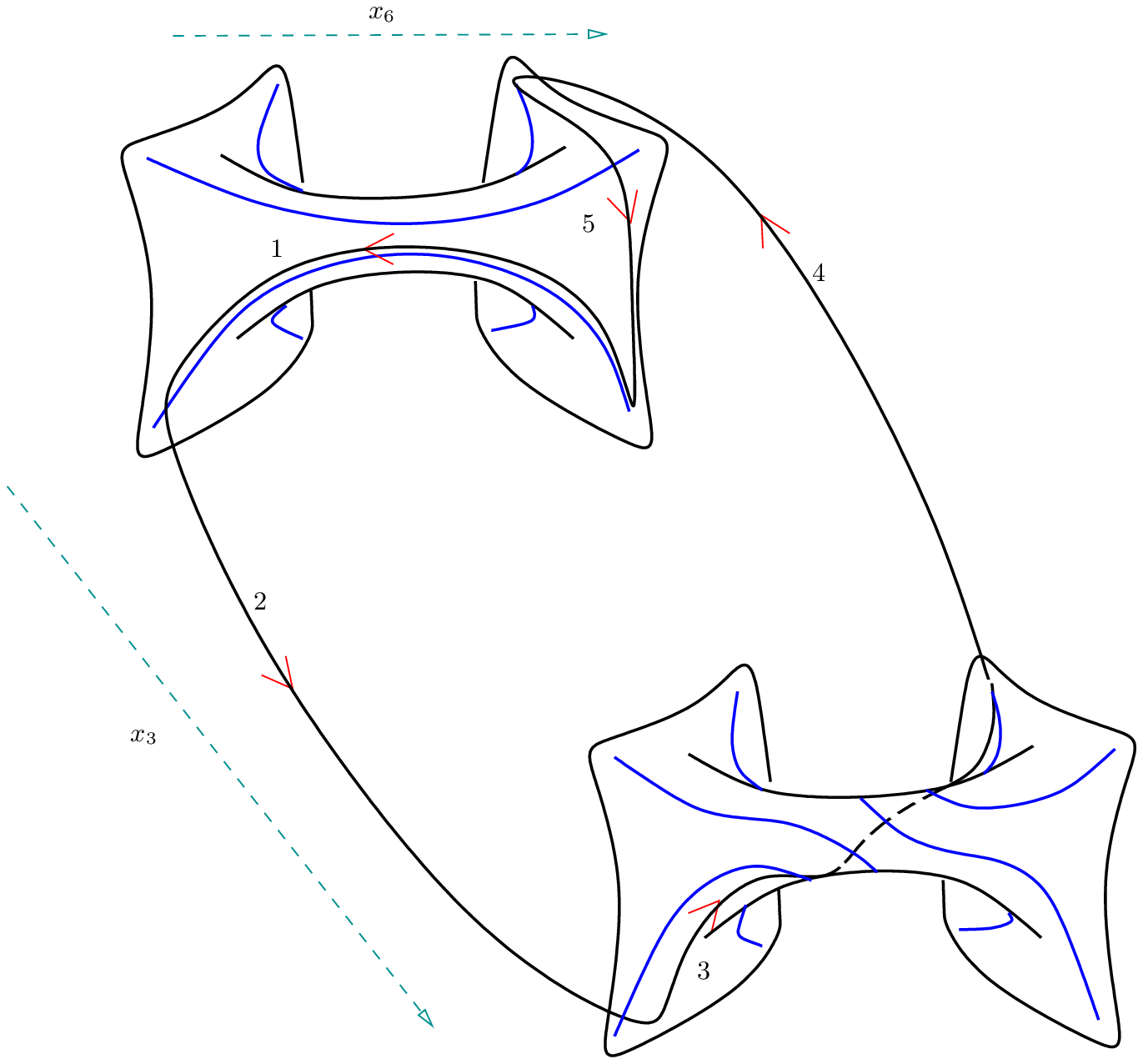}
\caption{An example of domain wall $S_2$
interpolating between the surfaces $\Sigma_0$ at $x_3 \to -\infty$
and $\Sigma_2$ at $x_3 \to +\infty$ in SU$(4)$ SYM. $S_2$ is a three
dimensional manifold with two non-trivial cycles. One of them,
peculiar to the domain wall, is described in the figure as a
composition of five pieces. It winds twice around the M-theory
circle. }
\label{wall}
}
It thus implies that $\Z_{[k,n]}$ strings are stable in the $k$-wall
background. If perpendicular to the wall, the $\Z_{[k,n]}$ strings
can cross it without changing their $[k,n]$ quantum number. If
parallel to the wall they shall set in the most energetically
favorable $x^3$ position. For a detailed study we should consider
the elements of $H_1(\widetilde{Y}/S_k,\Z)$ and minimize their
length. This would certainly require a more detailed understanding of
the manifold $S_k$ (some results can be found in \cite{Volo}).

Another point to discuss is the topology. One may, in fact, suspect
that the topology of the manifold $S_k$ could have some other
non-trivial cycles that could ruin our homology group computation.
We think this is not the case. The reason is the following. If we
make a large $n$ limit keeping $k/n$ fixed, we expect the
domain wall to behave as a soliton of effective Lagrangian of the kind \beq
{\cal L}_{eff} \propto n^2 F(\phi,\vec\nabla \phi, \dots) \ . \eeq So
its tension should scale like $n^2$ and, most importantly, its spatial
dependence and size, being determined only by ${\cal F}$ should be
$n$ independent. So for example, if we take a $k$-wall in a SU$(n)$
gauge theory, it should have the same profile of a $k/[k,n]$-wall in
a SU$(n/[k,n])$ gauge theory. The only difference is in their
tension whose ratio is $[k,n]^2$. In MQCD we expect the manifold
$S_k$ of the SU$(n)$ theory, to have the same spatial properties,
and in particular the same topology, of the manifold $S_{k/[k,n]}$
in the SU$(k/[k,n])$. The only difference is that the former winds
$[k,n]$ times more around the M-theory circle. So the topology is
not changed.

%% file: gaugegravity.tex
Gauge-gravity correspondence relates a certain gauge theory to a
string theory in a particular background. We now consider the
gauge-gravity realization of $\n=1$ SYM due to Maldacena-Nunez
\cite{MN} and Klebanov-Strassler \cite{KS} solutions. The differences between
these two realizations are not relevant for what we are going to
say. What is important for us is the chiral symmetry breaking at the
end of the cascade. We begin with a brief summary of this result,
then introduce the $k$-wall in this framework and finally explain the
emergence of $\Z_{[k,n]}$ strings.

The string theory under consideration is Type IIB. Space-time is
composed by the $3+1$ dimensional space time where the gauge theory
lives, times a warped dimension corresponding to the energy scale of
the gauge theory, times an internal manifold of topology $S^2 \times
S^3$. $n$ units of $F_3^{\rm RR}$ flux pass through the $S^3$ sphere.
Confining strings are F$1$-strings and can be annihilated in units
of $n$ ending on a baryon vertex. The baryon vertex is a D$3$-brane
wrapped on the $S^3$ sphere. A Chern-Simons interaction with the
$F_3^{\rm RR}$ form requires $n$ fundamental strings to end on the
baryon vertex \cite{wittenbaryons}. A domain wall is a D$5$-brane
wrapping the $S^3$ sphere. The $1$-strings ending on a $1$-wall is
now an F$1$-string ending on the D$5$-brane.

\FIGURE[h!t]{
\includegraphics[width=16em]{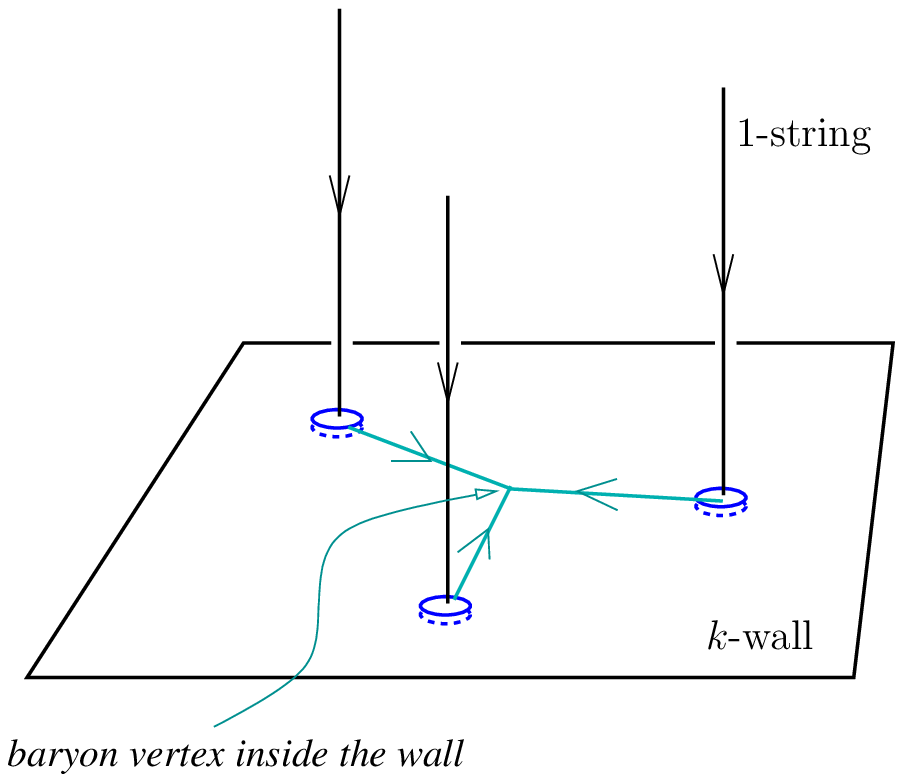}
\caption{A baryon vertex living on the
domain wall provides the mechanism for  $[k,n]$ separate $1$-strings
to end on the wall.}
\label{baryonvertex}
}
We want now to consider the $k$-wall that consists of $k$
D$5$-branes superimposed. The phenomenon we want to see, that is the
$\Z_{[k,n]}$ strings bounded inside the domain wall, is a non-perturbative effect from the point of view of the domain wall
effective action. The way to see it is to consider a large number of
domain walls ($k,n \to \infty$ while keeping $k,n$ fixed) and
consider the 't Hooft limit of this low energy effective action. We
should thus consider the black brane description of the $k$
D$5$-branes and go to the near-horizon geometry. Clearly, this is a
difficult task, we do not even know the solution of the black
$5$-brane in this fields' background. But our goal requires much less
than the full solution of the gauge-gravity dual of the $k$-wall
effective action. We just need to know that the horizon of these $k$
D$5$-branes is a three dimensional manifold, with topology $S^3$,
and with $k$ units of $F_3^{\rm RR}$ flux passing through it. A
D$3$-brane wrapping this manifold is a kind of baryon vertex for
this $2+1$ effective theory. The Chern-Simons interaction with the
$F_3^{\rm RR}$ flux requires $k$ strings to end on it.

We thus have found two baryon vertices on which strings can
annihilate. One is that of the original gauge theory, living in
$3+1$ dimensions, on which $n$ F$1$-strings can end. The other is
the baryon vertex of the $k$-wall that lives on its $2+1$ world
volume. The last can annihilate $k$ units of F$1$'s. Strings can
thus be annihilated in any integral linear combinations of $n$ and
$k$, and so the $\Z_{[k,n]}$ stability follows.

The baryon vertex living on the $k$-wall solves an apparent puzzle of our previous analysis. We said, in fact, that a $1$-string can not end on a $k$-wall (consider $[k,n] =k$ for
simplicity here) but a $k$-string does. But what happens if we take
$k$ $1$-strings? They should be able to terminate on the $k$-wall
although the single $1$-strings are not able to. What happens is explained
by the $k$-wall baryon vertex (see Figure \ref{baryonvertex}). From
every end point of $1$-strings, a string inside the wall departs.
These $k$ strings are then annihilated into a $k$-wall baryon
vertex.

%{\bf What about the back reaction? Is there a geometric transition?}
In the previous discussion we have neglected the back-reaction of the D$3$-branes on the geometry. This certainly must be taken into account for a full treatment of the problem. We can nevertheless consider the limit $k$ fixed, $[k,n]$ fixed and $n \to \infty$. In this limit we can neglect the back-reaction of the $k$ D$3$-branes, and the previous discussion can be considered a valid support of our main statement.

%% file: fieldtheory.tex
The string inside wall phenomenon has been studied in detail in  the paper \cite{Auzzi:2008zd}, following the initial idea of \cite{Dvali:2007nm}.  Strings can form a bound state with a domain wall if there is a charged condensate that \textit{does not} vanish on both sides of the wall. This basic mechanism is also the one responsible for the string-wall bound state in $\N=1$ SYM.
Let us consider the
$k$-wall that interpolates between the $0$ and the $k$ vacua, where
the two condensates responsible for confinement are, respectively, the monopole
$(0,1)$ and the dyon $(k,1)$. Strings inside walls are formed by the tunneling of the condensed particles, from one vacuum to the other. This tunneling  is possible only if $k$ and $n$ have some divisor in common. We can, in fact, take a bunch of $n/[k,n]$ monopoles one one side and make them reappear on the other side as $n/[k,n]$ dyons. The charges match since we have  $(0,n/[k,n])$ on the monopole side and $(n/[k,n] \cdot k ,n/[k,n])=(0,n/[k,n])$ on the dyon side.  This tunneling is responsible for the formation of the $[k,n]$ confining strings inside the $k$-wall in SU$(n)$ super Yang-Mills.

$\N =1$ super Yang-Mills is still far from being under complete analytical control.  The fact that it lies in a non-Abelian kind of confining phase, at the base of the heuristic argument,  has not yet a direct and rigorous proof.  To have a more solid field theoretical setting we shall investigate in what follows the soflty broken $\N=2$ and detect here the string inside wall phenomenon.

\vskip 0.50cm
\begin{center}
*  *  *
\end{center} 
\noindent

We consider  the deformation from $\N=2$ through a mass term $\mu \Tr \Phi^2 /2$ for the adjoint chiral superfield. 
The theory for $\mu \ll \Lambda_{\snd}$ is under analytical control thanks to the Seiberg-Witten solution. The gauge group abelianizes $\SU(n) \to \U(1)^{n-1}$ and each $\U(1)$ is Higgses by the condensation of an opportune low-energy hyper-multiplet. Duality between the microscopic and macroscopic description implies confinement of the original electric probe charges. This phase still persists as we change the mass $\mu$. In particular, the $n$ vacua are continuously deformed in the $n$ vacua of pure $\N=1$ as $\mu \gg \Lambda_{\snu}$. An important difference though, is that there is no abelianization in the $\mu \to \infty$ limit and the nature of the confining phase is purely non-Abelian.

We now want to try the deformed $\N=2$ technique to understand the phenomenon in which we are interested in this paper. The basic prototype of the domain wall in deformed $\N=2$ has been studied in Ref.~\cite{Kaplunovsky:1998vt}, at least for the simplest case $n=2$. 
The SW curve for $\SU(n)$ $\N=2$ SYM is:
\bea
\label{}
{y}^2 &=& {\cal P}_{n}(z) \nonumber \\
&=& \frac{1}{4} \det(z-\phi)^2-  \Lambda^{2 n} 
\eea
where we defined for convenience, the polynomial ${\cal P}_{n}$.
The maximal singularity points for ${\cal P}_{n}$ are given by the solution 
of Douglas and Shenker  \cite{DS}. There are $n$ of these maximal singularity points. They happen when the $n$ cuts are lined up and all the roots, a part from two of them, are doubled. One solution is when all the roots are on the real axis. The others are related by an $e^{2\pi i k / n}$ transformation.

In the real case we can take $\phi = {\rm diag} (\phi_1,\dots,\phi_{n})$ and $\phi_j = 2\Lambda  \cos{\left( \pi \frac{ j-1/2}{n}\right)} $ and the curve is thus written in terms of  Chebyshev functions:
\bea
{\cal P}_{n } &=& \frac{1}{4} \prod_{j=1}^n \left(z - 2 \Lambda\cos{\left( \pi \frac{ j-1/2}{n}\right)} \right)^2 - \Lambda^{n } \nonumber \\
&=& \left( \frac{1}{4} T_{n}\left(\frac{z}{2 \Lambda }\right)^2 - 1\right) \Lambda^{n} \nonumber \\
&=&   \left(\frac{z^2}{4} -\Lambda^2 \right)  U_{n-1}\left(\frac{z}{2 \Lambda}\right)^2   \Lambda^{n-2}
\label{real}
\eea
where $ U_{n-1}\left(\frac{z}{2 \Lambda}\right)^2 = \prod_{j=1}^{n -1} \left( \frac{z}{2 \Lambda} -  \cos{  \frac{\pi  j}{n}} \right) $. For the factorization of the curve  we have used the important identity:
\beq
T_n ^2 (z) -(z^2 -1)U_{n-1}(z) = 1 \ .
\eeq
We recall what  the relation is between the curve and the $\Z_n$ strings. 
Massless particles appear every time there is a vanishing cycle in the SW curve. In the maximal singularity vacua there are $n-1$ double roots, and so $n-1$ massless hypermultiplets, one for every $\U(1)$ factor in the low-energy effective action.
 Upon the perturbation with the mass term $\mu$, these massless particles condensed and creates vortices (ordinary Abrikosov-Nielsen-Olesen vortices in the low-energy effective action).  These $n-1$ vortices are exactly in one-to-one correspondence with the non-trivial elements of the group of confining $\Z_n$ strings \cite{DS,HSZ,Bolognesi:2004da}:
\bea
{\cal T}_k &=& 4\pi \tE_k E_k  \nonumber \\
&=&  4\pi  \left.\sqrt{W^{\prime 2}(z) + f(z)}\right|_{z=2\Lambda \cos{(\pi k /n)}}  \nonumber \\
&=& 8\pi \mu \Lambda \sin{\frac{\pi k}{n}} \ ,
\eea
where we have  derived the Douglas-Shenker sine formula for the k-string tension.

For the effect we are interested in, the simplest case to consider is $n=4$ and $k=2$, that is, the gauge group $\SU(4)$ and the $2$-wall. We choose the $2$-wall to interpolate between the $h=0$ vacuum, the real one, and the $h=2$ vacuum, the imaginary one. Since $[k,n]=2$, we expect the $1$-string to be stable in the $2$-wall background; a parallel $1$-string from a bound state inside the $2$-wall. The $2$-string can instead terminate on the $2$-wall, exactly like the $1$-string terminates on the $1$-wall for the gauge group $\SU(2)$.

It is better to start with the $\SU(2)$ case that is a very well known example, and it also appears additionally as a substructure of the $\SU(4)$ problem we shall face next. Now the SW curve has degree four and genus one. There is one $\U(1)$ gauge group in the low-energy theory. There are two vacua of maximal singularity, one real and one imaginary.  In Figures \ref{ciclidue} and \ref{vanishingciclidue} we have the roots of the SW curve near the real vacuum. The first has the cycles $\alpha$ and $\beta$ corresponding to the electric and magnetic components of the unbroken $\U(1)$. The second has the vanishing cycle of the particle that becomes massless in the imaginary vacuum. 
\DOUBLEFIGURE{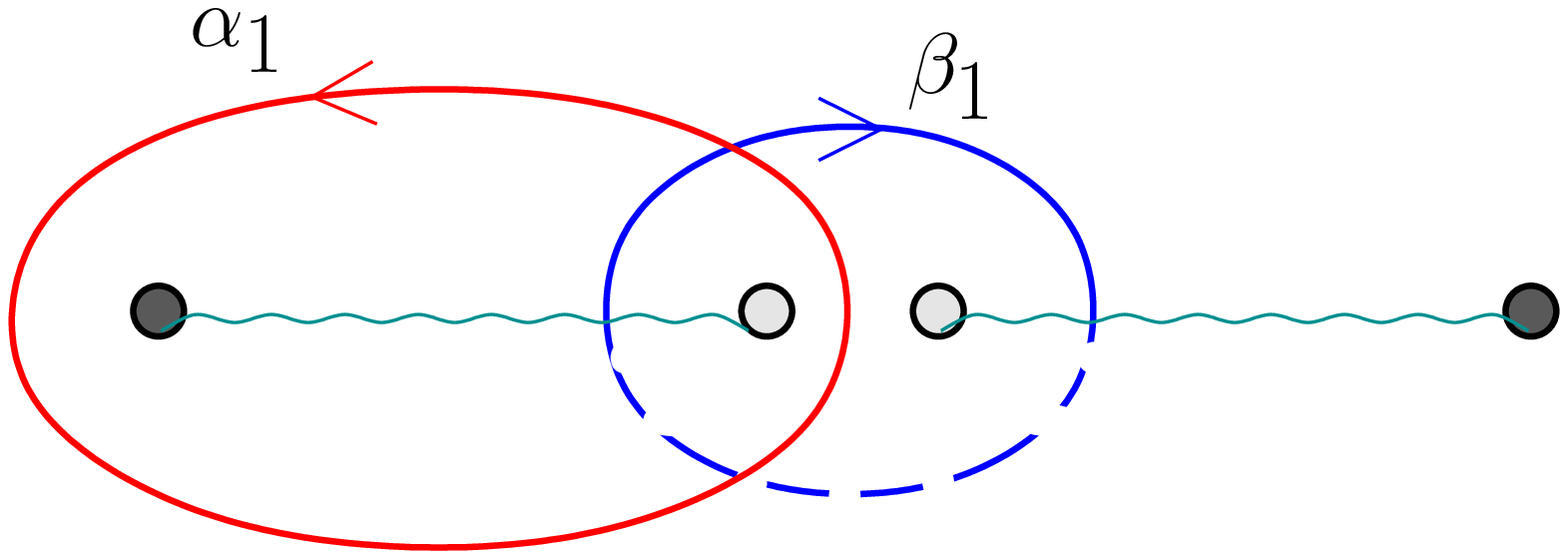,width=15.5em}{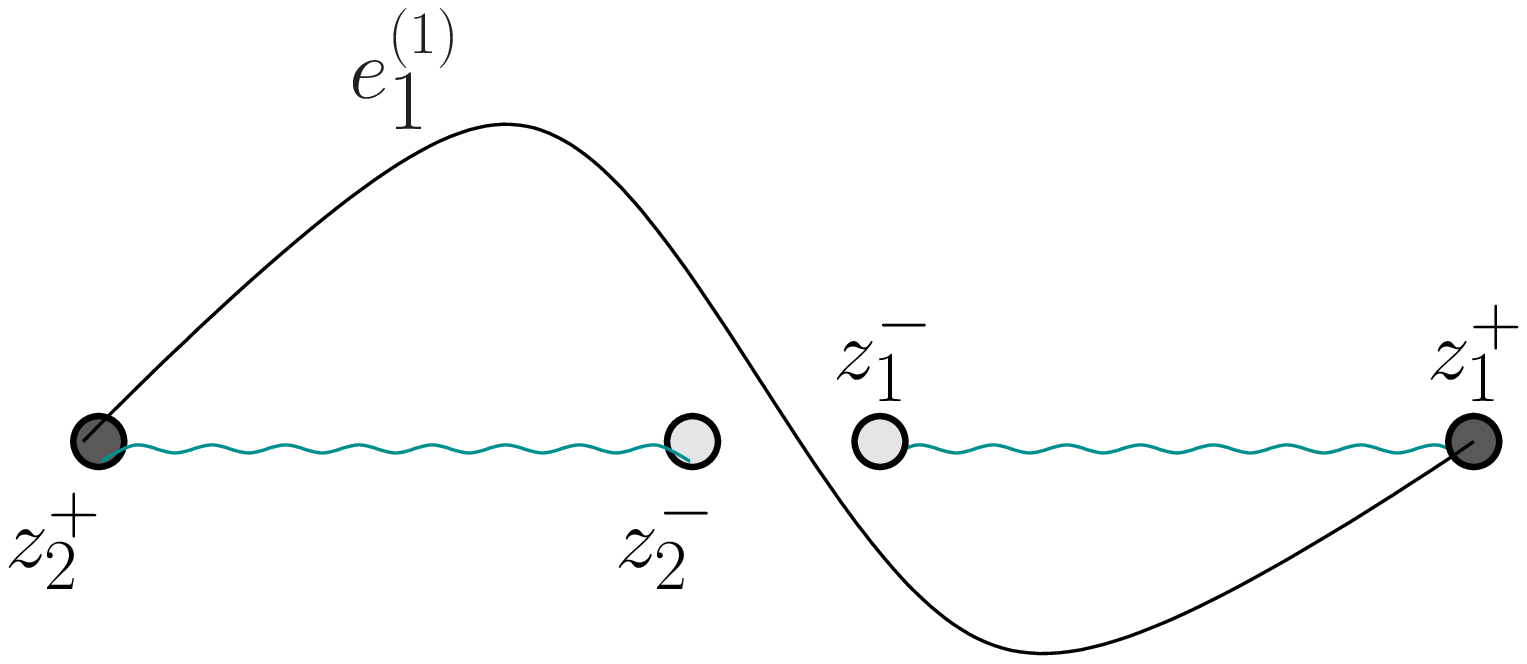,width=15.5em}{\label{ciclidue} Roots and cycles near the real vacuum \ref{real}. The choice of basis is made so that the $\alpha$ corresponds to the electric in weak coupling.}{\label{vanishingciclidue} The same roots but now with the vanishing cycles $E^{\1}$ corresponding to the roots that collide in the dyon vacuum.}
In the real vacuum the particle that become massless is $E^{\0}$ with charges $(0,1)$ with respect to the $\U(1)$ (the first is the electric charge and the second the magnetic charges). The particle $E^{\1}$ that is massless in the imaginary vacuum has charges $(2,-1)$.
We know that the $1$-string (the only string in this case) is an ANO vortex created by $E^{\0}$. Through the screening mechanism it can terminate on the $1$-wall. The condition for this to be possible is that the flux carried by the string  can be expressed as the sum of the charged particles that condense on both sides of the wall:\footnote{We use $E$ to denote the fields and $e$ to denote the charges and/or the cycles.}
\bea
\label{termino}
{\rm flux} &=&  \nu^{\0}  e^{\0} + \nu^{\1}  e^{\1} \ ,
\eea
This condition  is solved for the $1$-string carrying flux $(1,0)$ by:
\bea
 \nu^{\0} = 1/2 \ , \qquad  \nu^{\1} = -1/2 \ .
\eea
We shall discuss more in detail this string-wall junction, and how to quantitatively approach the problem, in Appendix \ref{junction}.

Now let us move to the more interesting $\SU(4)$ case.
The curve has $8$ roots and they can be divided into {\it plus} and {\it minus} roots according to the following factorization:
\bea
{\cal P}_{4 } &=& \frac{1}{4}  \det(z-\phi)^2 - \Lambda^8  \nonumber \\
&=&  \left( \frac{1}{2}   \det(z-\phi) - \Lambda^4 \right) \left( \frac{1}{2}  \det(z-\phi)  + \Lambda^4 \right) \nonumber \\
&=&  {\cal P}_{4}^{-} {\cal P}_{4}^{+}
\eea
We then give names to the various roots: 
\beq
z^{-}_{1,2,3,4} \ , \qquad z^{+}_{1,2,3,4} \ .  
\eeq
We want to focus our attention on the two vacua $h=0$ and $h=2$, respectively, the real  and the imaginary vacua. In  the real case the factorization gives the following roots: 
\bea
 {\cal P}_{4}^{-} &=&  \frac{1}{2}  z^2 (z-2\Lambda)(z+2\Lambda) \nonumber \\
{\cal P}_{4}^{+}  &=&   \frac{1}{2} (z-\sqrt{2}\Lambda)^2 (z+\sqrt{2}\Lambda)^2 \ ,
\eea
In the imaginary case:
\bea
 {\cal P}_{4}^{-} &=&  \frac{1}{2}  z^2 (z-2i\Lambda)(z+2i\Lambda) \nonumber \\
{\cal P}_{4}^{+}  &=&   \frac{1}{2} (z-\sqrt{2}i\Lambda)^2 (z+\sqrt{2}i\Lambda)^2 \ .
\eea
Six of the roots are paired, and as a consequence each of the $\U(1)^3$ low-energy gauge groups has a massless charged hypermultiplet.
The roots shuffling is important for our purposes.  To compute the charges of the massless particles in a given vacuum we need the vanishing cycles corresponding to the given massless particles. To get them is not only necessary to know which roots collide, but also to understand the path they follow with respect to the other roots.

\DOUBLEFIGURE{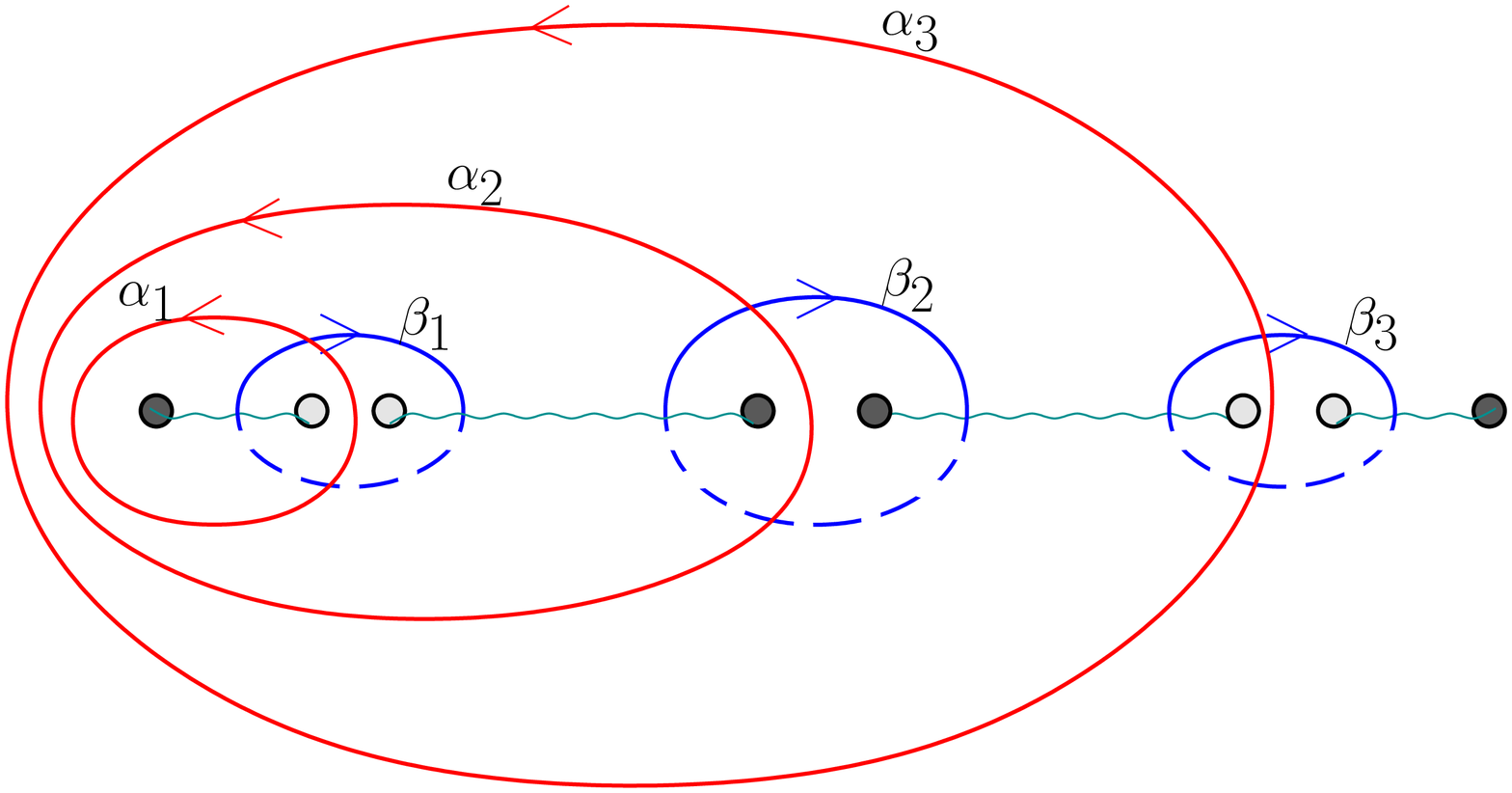,width=19.5em}{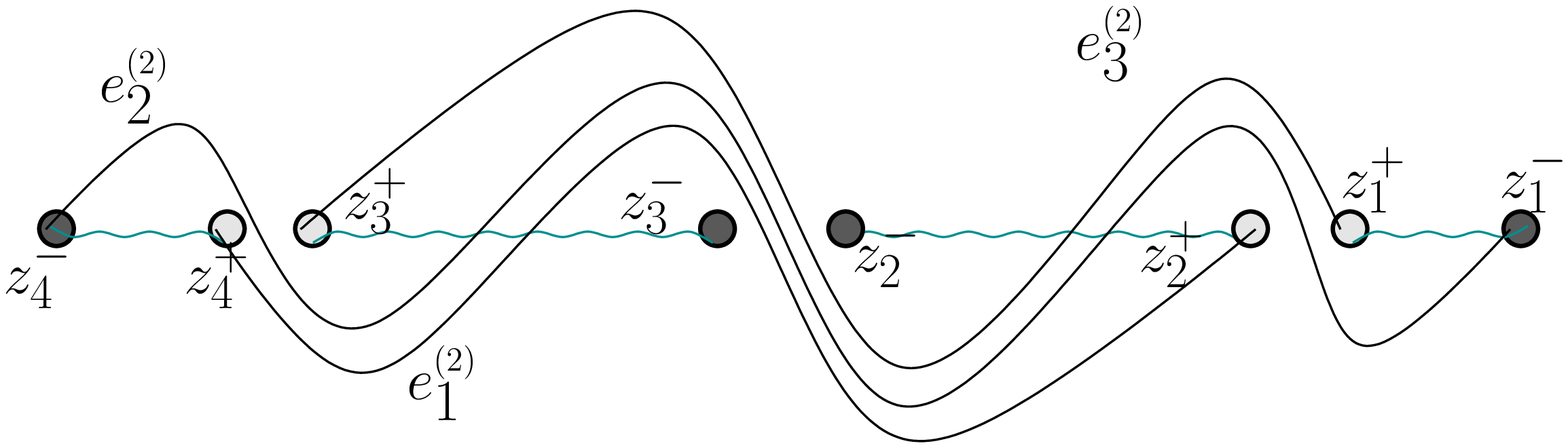,width=19.5em}{\label{cicli} Roots and cycles near the real vacuum \ref{real}. The choice of basis is made so that the charges of massless particles are diagonal (Table \ref{chargesreal}).}{\label{vanishingcicli} The same roots, but now with the vanishing cycles $e_1^{\2},e_2^{\2},e_3^{\2}$ corresponding to the imaginary vacuum ((Table \ref{chargesim})).}
Passing from the vacuum $h=0$ to the vacuum $h=0$, the two set of roots $z^{-}_{1,2,3,4}$ and $z^{+}_{1,2,3,4}$  are shuffled independently. The roots  $z^{+}_{1,2,3,4}$ are the one responsible for the massless particles $E_1^{\2}$, $E_3^{\2}$ and $E_1^{\0}$, $E_3^{\0}$.
In this particular case, the condensation of $E_1$ creates the $1$-string, condensation of $E_2$ the $2$-string and condensation of $E_3$ the $3$-string of $\Z_4$.

We then use the \textit{same} basis to compute the charges for the imaginary vacuum (Table \ref{chargesim}). Of course we could have used another basis in which the charges would have locked diagonal exactly like in the real vacuum. But since we want to study the domain wall between the two vacua, we need a description in which the three $\U(1)$s are expressed on the same basis in the two vacua. In Figure \ref{vanishingcicli} we displayed the vanishing cycles corresponding to the massless particles in the imaginary vacuum. Expanding these cycles on the basis previously given gives us the charges of Table \ref{chargesim}.
\DOUBLETABLE[h]{\begin{tabular}{c|ccc}
& $\U(1)_1$   & $\U(1)_2$   &  $\U(1)_3$ \\ \hline
$e_1^{\0}$  & $(0_{\rm e},1_{\rm m})$   &    & \\
$e_2^{\0}$  &       &   $(0_{\rm e},1_{\rm m})$   & \\
$e_3^{\0}$  & &   & $(0_{\rm e},1_{\rm m})$ \\ 
\hline
\end{tabular}
}{\begin{tabular}{c|ccc}
& $\U(1)_1$   & $\U(1)_2$   &  $\U(1)_3$ \\ \hline
$e_1^{\2}$  & $(-1_{\rm e},1_{\rm m})$   & $(2_{\rm e},-1_{\rm m})$    &$(-1_{\rm e},0_{\rm m})$ \\
$e_2^{\2}$  &  $(-2_{\rm e},1_{\rm m})$       & $(2_{\rm e},-1_{\rm m})$   & $(-2_{\rm e},1_{\rm m})$  \\
$e_3^{\2}$  & $(-1_{\rm e},0_{\rm m})$ & $(2_{\rm e},-1_{\rm m})$  & $(-1_{\rm e},1_{\rm m})$ \\ 
\hline
\end{tabular} 
}{ \label{chargesreal}Low-energy gauge groups and charged hypermultiplets for the real vacuum (\ref{real}), $n=4$, $h=0$.}{\label{chargesim}Low-energy gauge groups and charged hypermultiplets for the imaginary vacuum (\ref{real}), $n=4$, $h=2$. We have used the same basis of Table \ref{chargesreal}. }

The screening condition (the generalization of (\ref{termino}))  is now:
\bea
{\rm flux} &=&  \nu_1^{\2}  e_1^{\2} + \nu_2^{\2}  e_2^{\2}+ \nu_3^{\2}  e_3^{\2}  \nonumber \\
&=& \nu_1^{\0} e_1^{\0} + \nu_2^{\0} e_2^{\0} + \nu_3^{\0} e_3^{\0}  \label{termino4}
\eea
which is solved for the $2$-string carrying flux $(0,0),(1,0),(0,0)$ by:
\bea
 \nu_1^{\2} = 1/2 \qquad  \nu_2^{\2} = -1/2 \qquad\nu_3^{\2} = 1/2 \nonumber \\
\nu_1^{\0} =0 \qquad  \nu_2^{\0} = -1/2 \qquad  \nu_3^{\0} = 0
\eea
Note that we have a redundancy, since two equations (the electric of $\U(1)_1$ and $\U(1)_3$) are equivalent: $0=\alpha+2\beta-\gamma$.
This is related to the reason why the vortices created by $E_1^{\0}$ and $E_1^{\0}$ can not be {\it screened}. Consider the $1$-string with flux $(1,0),(0,0),(0,0)$. The electric of $\U(1)_1$ and $\U(1)_3$) are now, respectively,  $1=\alpha+2\beta-\gamma$ and $0=\alpha+2\beta-\gamma$; clearly there is no solution to the termination condition (\ref{termino4}) in this case.

We note also that the vectorial space spanned by the composite condensate $\widetilde{E}_1^{\0} E_3^{\0}$ is exactly the same as that spanned by $\widetilde{E}_1^{\2} E_3^{\2}$.
The product of the fields implies the sum of the charges: 
\beq
-e_1^{\0} +e_3^{\0} = - e_1^{\2} + e_3^{\2} \ .
\eeq
 So the situation is completely analogous to the one described in \cite{Auzzi:2008zd}. The $1$-string and the $3$-string can not terminate on the wall.  If parallel to it, they are attracted to form a bound state consisting of a confining string inside the domain wall. The lowest energy configuration for the string is where the condensates reaches their minimum, and that is in the middle of the wall.

We can also analyse the SU$(6)$ gauge theory. As before, we consider a domain wall with the $h=0$ vacuum on on e side.  We again choose the basis of cycles so that the charges appears simple in this vacuum. There are five $\U(1)_i$ gauge groups and five hyper-multiplets $E^{\0}_i$ with diagonal charges  $(0_{\rm e},1_{\rm m})$.  
With $n=6$ there are two interesting domain walls we can consider.  One is the $3$-wall between the $0$ vacuum and the $3$ vacuum. Keeping the same base choice as before, the chrages of the massless particles in the $2$-vacuum are given in Table \ref{tresei}.
\TABLE{
\begin{tabular}{c|ccccc}
& $\U(1)_1$ & $\U(1)_2$ & $\U(1)_3$ & $\U(1)_4$ &  $\U(1)_5$ \\ \hline
$e_1^{\3}$  & $(0_{\rm e},0_{\rm m})$       & $(-1_{\rm e},0_{\rm m})$ & $(2_{\rm e},-1_{\rm m})$  & $(-1_{\rm e},1_{\rm m})$ &$(0_{\rm e},0_{\rm m})$ \\
$e_2^{\3}$  &  $(1_{\rm e},0_{\rm m})$      & $(-2_{\rm e},1_{\rm m})$ & $(2_{\rm e},-1_{\rm m})$  &$(-2_{\rm e},1_{\rm m})$ & $(1_{\rm e},-1_{\rm m})$  \\
$e_3^{\3}$  & $(2_{\rm e},-1_{\rm m})$      &$(-2_{\rm e},1_{\rm m})$ & $(2_{\rm e},-1_{\rm m})$ &$(-2_{\rm e},1_{\rm m})$ & $(2_{\rm e},-1_{\rm m})$ \\ 
$e_4^{\3}$  & $(1_{\rm e},-1_{\rm m})$      &$(-2_{\rm e},1_{\rm m})$  & $(2_{\rm e},-1_{\rm m})$  & $(-2_{\rm e},1_{\rm m})$ &$(1_{\rm e},0_{\rm m})$ \\
$e_5^{\3}$  &  $(0_{\rm e},0_{\rm m})$      & $(-1_{\rm e},1_{\rm m})$ & $(2_{\rm e},-1_{\rm m})$  & $(-1_{\rm e},0_{\rm m})$& $(0_{\rm e},0_{\rm m})$  \\
\hline
\end{tabular} \label{tresei} \caption{Low-energy gauge groups and charged hypermultiplets for the  $n=6$, $h=3$.}
}
Since $[k,n]$ is now equal to $3$, we expect the $3$-string to be able to terminate on the domain wall. The screening condition, as we can see:
\bea
3-{\rm string} &=& (0,0);(0,0);(1,0);(0,0);(0,0)  \nonumber \\
&=&
\frac12\left( e_3^{\3} +e_3^{\0} -e_2^{\3}-e_4^{\3}+e_1^{\3}+e_5^{\3}  \right) \ ,
\eea
is in fact solvable. The $1$-string and $2$-string are, instead, stable. The reason is the equivalence between the following equivalences of the condensates:
\bea 
e_1^{\3} - e_5^{\3} &=& -e_2^{\0}+e_4^{\0}  \ , \nonumber \\
 -e_2^{\3} + e_4^{\3} &=& -e_1^{\0} + e_5^{\0} \ .
\eea

We can complete the discussion with the $2$-wall in the SU$(6)$ theory. Charges in the $2$-vacuum are now given in Table \ref{duesei}.
\TABLE{
\begin{tabular}{c|ccccc}
& $\U(1)_1$ & $\U(1)_2$ & $\U(1)_3$ & $\U(1)_4$ &  $\U(1)_5$ \\ \hline
$e_1^{\2}$  & $(0_{\rm e},0_{\rm m})$ & $(0_{\rm e},0_{\rm m})$ & $(1_{\rm e},0_{\rm m})$  & $(-2_{\rm e},1_{\rm m})$ &$(1_{\rm e},-1_{\rm m})$ \\
$e_2^{\2}$  &  $(0_{\rm e},0_{\rm m})$      & $(-1_{\rm e},0_{\rm m})$ & $(2_{\rm e},-1_{\rm m})$  &$(-2_{\rm e},1_{\rm m})$ & $(2_{\rm e},-1_{\rm m})$  \\
$e_3^{\2}$  & $(1_{\rm e},0_{\rm m})$&$(-2_{\rm e},1_{\rm m})$ & $(2_{\rm e},-1_{\rm m})$ &$(-2_{\rm e},1_{\rm m})$ & $(1_{\rm e},0_{\rm m})$ \\ 
$e_4^{\2}$  & $(2_{\rm e},-1_{\rm m})$ &$(-2_{\rm e},1_{\rm m})$  & $(2_{\rm e},-1_{\rm m})$  & $(-1_{\rm e},0_{\rm m})$ &$(0_{\rm e},0_{\rm m})$ \\
$e_5^{\2}$  &  $(1_{\rm e},-1_{\rm m})$      & $(-2_{\rm e},1_{\rm m})$ & $(1_{\rm e},0_{\rm m})$  & $(0_{\rm e},0_{\rm m})$& $(0_{\rm e},0_{\rm m})$  \\
\hline
\end{tabular} \label{duesei} \caption{Low-energy gauge groups and charged hypermultiplets for the  $n=6$, $h=2$. }
}
The $2$-string can terminate on the domain wall. The screening condition is:
\bea
2-{\rm string} &=& (0,0);(1,0);(0,0);(0,0);(0,0)  \nonumber \\
&=&
\frac13\left( -e_2^{\2} + 2 e_3^{\2} -2e_4^{\2} + 2e_5^{\2} \right. \nonumber\\
 && \left. -2e_2^{\0}-e_3^{\0}-e_4^{\0}-e_5^{\0}  \right) \ .
\eea
The $1$-string is stable due to the following equality
\beq
- e_1^{\2} +   e_3^{\2} -  e_5^{\2} = 
e_1^{\0} -  e_3^{\0}+ e_5^{\0} \ .
\eeq

%% file: objection.tex
The reader has probably an objection in his mind. 
Here we want to promptly discuss and solve this apparent puzzle.

According to the previous statement, a $1$-string can not terminate to a $2$-wall in the case where
the number of colors $n$ is even (the simplest example of
$[k,n]$ non-trivial effect). But there could be a {\it peeling} of
the $2$-wall. A
condensate with the required charge could thus be formed near the first peeled sheet, that corresponds to a $1$-wall. The peeling seems to require a
certain finite amount of energy, but this would be gained by the fact
that we do not need another string at the opposite side of the wall.
It appears that in this way we gain an infinite amount of energy at
the price of just a peeling of a small portion of the $2$-wall.

This objection, as we expressed it, contains two
mistakes.

First of all, the energy difference between a string ending on a wall
and a string crossing the wall and proceeding on the other half
space is not infinite, but finite. The reason lies behind the
logarithmical bending that a string produces on a wall where it
terminates.  When a string
of tension $T_\S$ ends on a wall of tension $T_\W$ it produces a
deformation of the profile of the wall. At large distance this
deformation is $f(r)=( T_\S / \pi T_\W) \log{r}$. To evaluate the
energy of the wall we perform a surface integral $T_\W \int 2\pi r dr
\sqrt{1+f^{\prime 2}}$. At large distance the derivative goes to
zero and we can expand and we obtain two terms. The first corresponds
to the energy of a flat wall. The second term is $( T_\S ^2 / T_\W)
\int^R dr 1/r$. Performing the integral we get $T_\S f(R)$,
that is exactly the energy of a string on the other side of a flat
wall. We thus see from this simple calculation that the energy of
the two configurations differs
only by a finite amount of energy: {\it the boojium}. We have thus
seen that there is no infinite amount of energy in the
game. A good understanding of this phenomenon can be found
in extended supersymmetric theories \cite{tongsakay,bojium} where
everything in BPS saturates and the boojium energy can be computed
exactly as a central charge.

\FIGURE[h!]{
\includegraphics[width=34em]{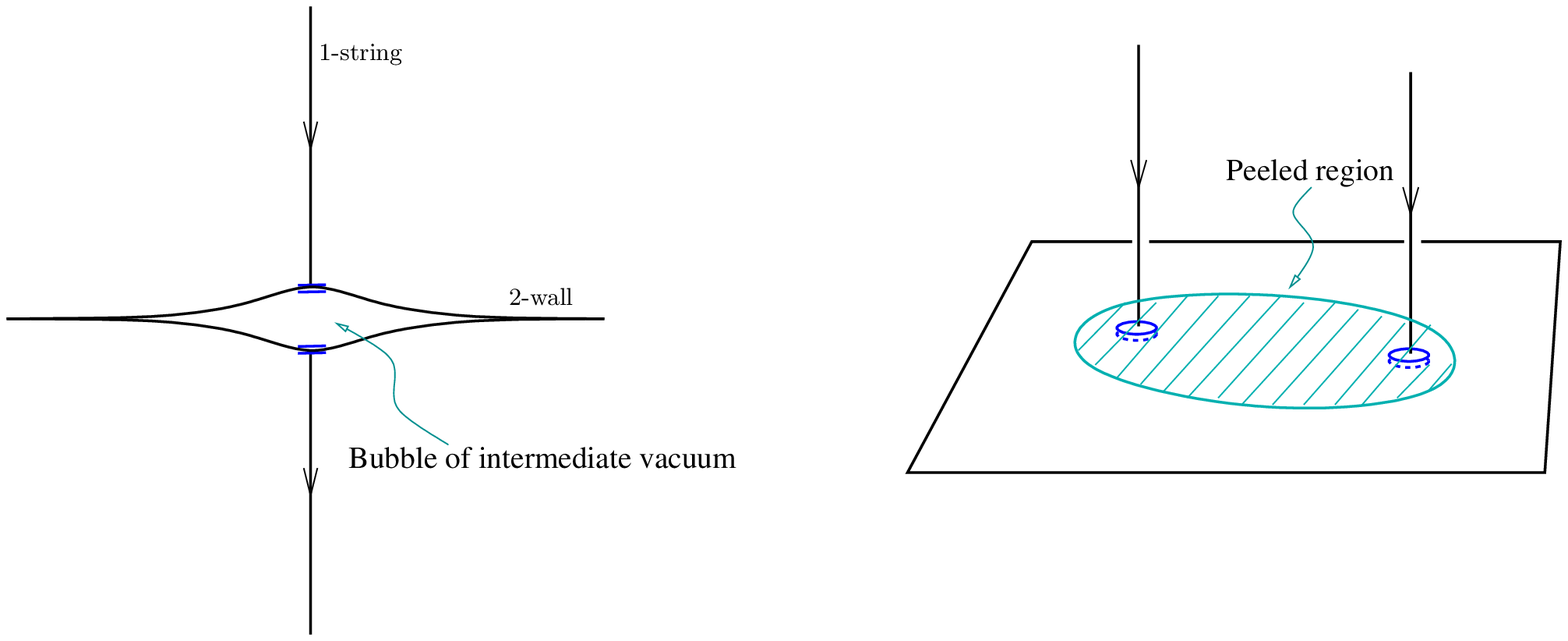}
\caption{The correct picture for $n$ even is
this one. When a bubble of intermediate vacuum is formed from the
peeling, the total charge of the particles inside the bubble must be
zero. }
\label{objectiondue}
}
But there is a more important reason why the objection does not stand. When we peel
the $2$-wall, the bubble of intermediate vacuum must have net zero
charge. So the correct picture is left of Figure \ref{objectiondue}.
This is just because we always have to satisfy a charge conservation
like (\ref{charge}).
We can move apart the two strings (right of Figure
\ref{objectiondue}), but there must always be a peeled region
connecting the two ending points.

This could be another intuitive
interpretation of the string that lies inside the wall. The string inside the wall is equivalent to a portion of peeled wall.

This fact can be further elucidated with another example that uses only well consolidated facts, without any relation to the string inside wall phenomenon. The two facts we are using are the following: {\it 1)} Confining strings, in particular the $1$-string, are stable objects in the vacuum; {\it 2)} A $1$-string can terminate on a $1$-wall. Now let's do the following experiment. We take a single $1$-string in the $0$-vacuum, and then create a bubble of the $1$-vacuum separated from the outside by a $1$-wall. 
\FIGURE[h!]{
\includegraphics[width=20em]{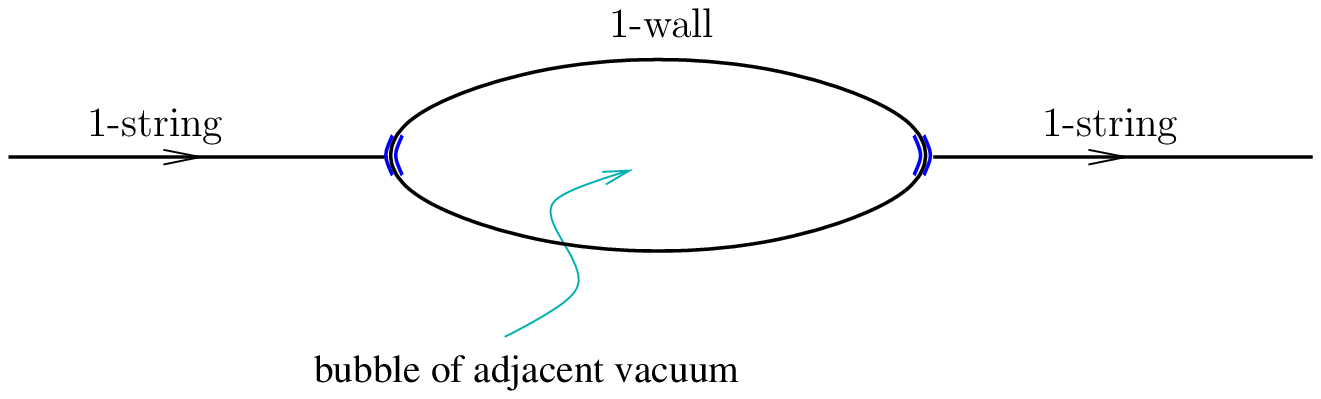}
\caption{A $1$-string  broken by a  $1$-wall bubble. }
\label{wellconsolidated}
}
So the $1$-string can be broken by the insertion of a bubble of $1$-wall containing inside an adjacent vacuum (the adjacent vacuum is inside the bubble). The bubble {\it can not} be broken and separated into two pieces. Otherwise we would be able to break the $1$-string into two disconnected pieces, like in a quark-antiquark formation. Being the $1$-string absolutely stable in this theory (there are no quark fields), we conclude that a vacuum bubble  must have total charge zero inside of it.  The argument just presented, made out of well consolidated facts, gives further understanding of the peeling phenomenon and the fact that the confining string is equivalent to a portion of peeled vacuum.

%% file: effectiveaction.tex
The effective action of the low-energy degrees of freedom of a
$k$-wall has been considered by Acharya-Vafa (AF) \cite{acharyavafa}. In the
string theory realization of $\n=1$ SYM previously considered, the
$k$ domain wall consists of $k$ D$5$-branes wrapped on a $S^3$
sphere. In space-time we get a $2+1$ brane with an $\N=2$ U$(k)$
gauge theory. The U$(k)$ gauge theory descends directly from the
gauge degrees of freedom on the branes. Here $\N=2$ is in $2+1$
dimensions, that is four real supercharges (like $\n=1$ in $3+1$
dimensions). We know that domain walls are half-BPS saturate, and so
we expect only two real supercharges corresponding to $\n=1$ in
$2+1$ dimensions. The breaking of supersymmetry derives from the
flux passing through the $S^3$ sphere. This induces an $\n=1$
Chern-Simon interaction at level $n$. The $2+1$ effective action, in
$\n=1$ language, consists thus of a gauge U$(k)$ multiplet with a
Yang-Mills and a Chern-Simons term coupled to an adjoint superfield.
Written explicitly in terms of the physical fields it is:
\begin{eqnarray}
&{\cal L}_{2+1}=& \frac{1}{g^2}\mathrm{Tr} \Big( - \frac{1}{2}
F_{\mu\nu} F^{\mu\nu}+ i \chi
D_{\mu}\gamma^{\mu} \chi  \nonumber \\[3mm] && \qquad
+ i \psi D_{\mu}\gamma^{\mu} \psi +  D_{\mu}\phi D^{\mu}\phi - \chi
[\phi, \psi] \Big)
 \nonumber \\[3mm]
&& +\frac{n}{4\pi}\mathrm{Tr} \Big( \frac{1}{2}
\epsilon^{\mu\nu\rho}(A_{\mu}
F_{\nu\rho}-\frac{2}{3}A_{\mu}A_{\nu}A_{\rho}) - \chi\chi \Big)
\end{eqnarray}
The fermion $\lambda$ is the supersymmetric partner of the gauge
field $A_{\mu}$ and acquires mass  through the supersymmetric
Chern-Simon term. The fermion $\psi$  is the supersymmetric partner
of $\phi$. Together, $A_{\mu}$, $\lambda$, $\chi$ and $\phi$ are the
field content of an $\n=2$ multiplet in $2+1$
dimensions\footnote{The first two lines of the Lagrangian are in
fact the dimensional reduction of $\n=1$ in $3+1$, and $\chi$,
$\psi$  are the two real components of the complex gaugino usually
denoted as $\lambda$.}. The Chern-Simons term splits their masses,
leaving only $\n=1$ residual supersymmetry.

The Acharya-Vafa theory has been derived in a string theory set-up. This setting is a parental to $\N=1$ SYM, but  not exactly the same.
Many questions about the  validity of the AF theory in the pure field theoretical context still remain unanswered. Every low-energy effective actions on domain walls has meaning and validity up to the scale of the inverse of the wall thickness. 
The theory has been put to a test only for the index \cite{acharyavafa,mishacounting} and succeeded in counting the expected number of vacua. But a more detailed understanding is certainly still needed.
In particular, is also not clear how to identify the gauge degrees of freedom of the Acharya-Vafa theory with the bulk ones. 
Moreover, we do not even  have a toy model that can reproduce the Chern-Simons term on the domain wall effective action.

With these warnings in mind, let us for the moment take the conservative approach that the Acharya-Vafa theory described part of the domain wall dynamics. We fist ask ourselves what  the phase of this theory is.

First of all consider the theory without the
Chern-Simons term. It is $\n=2$ in three dimensions. At tree level
there is a moduli space given by the expectation value of the scalar
field. From its holomorphic properties it is possible to compute the
non-perturbative generated superpotential \cite{AHW}. The result is
that there is a run-away vacuum and $\langle\phi\rangle$ goes to
infinity.  The Chern-Simons term breaks the $\n=2$ to $\n=1$ giving a
topological mass to the photon and to one real component of the
spinor. To higher loops in perturbation theory, the Chern-Simons term
generates a potential for the scalar field that stabilizes the
vacuum \cite{Armoni:2005sp}. From the fact that the domain wall is
BPS \cite{SD}, it is believed that the theory has a stable
supersymmetric vacua at a certain value of $\langle\phi\rangle = 0$.
This is consistent with what is known from the central charge of the
bulk theory that determines the tension of the walls.

Consider now the coupling dependence versus $n$; the coupling scales
like $1/g^2 \propto n/\Lambda$ where $\Lambda$ is
the dynamical scale of the original $\n=1$ theory in four
dimensions. There are two ways of performing a large $n$ limit. One is
to send $n$ to infinity while keeping $k/n$ fixed. This, in
fact, would be the proper large $k$ limit for the effective theory on
the domain wall. Another way is to send $n$ to infinity while
keeping $k$ fixed. This is the case we shall explore now. It is
simpler, since the theory on the domain wall maintains as fixed the gauge
group size and decreases the coupling.

For simplicity we shall thus restrict ourselves to a $2$-wall, that
is $k=2$, and take $n$ large. This is the simplest and most
tractable example to consider. The reason is that perturbation
theory is an expansion in powers of $g^2/\mu \propto
\Lambda/(n \mu)$ where $\mu$ is the energy scale. The theory
becomes strongly coupled at energies of order $\Lambda/n$.
The Chern-Simons term generates a topological mass for the photon of
order $m_{CS} = n g^2/(4\pi)  \propto \Lambda$ and thus where
the theory is weakly coupled. From this we can infer the following
conclusion. For arbitrarily large $n$ and fixed $k$, the theory is
weakly coupled at all scales.
The large $n$ analysis (with $k$ fixed) is thus under perturbative control. The theory is in the topological massive phase since the Chern-Simons mass acts at a scale where the gauge coupling is still small. More detailed analysis seems to confirm that this is the case for all $n$ and $k$ \cite{Witten:1999ds}.

This seems to go in the opposite direction of our claim. 
In the Acharya-Vafa theory there is no sign of confinement and strings inside the domain wall. To understand this point, and why this is not really in contradiction with our claim, we need to open a brief parenthesis on a supersymmetric toy model discussed in \cite{Auzzi:2008zd} which shares similar characteristics.

\vskip 0.50cm
\begin{center}
*  *  *
\end{center} 
\noindent
As already introduced in Section \ref{fieldtheory}, the purpose of Ref.~\cite{Auzzi:2008zd} is to understand, and find explicit realization of, the mechanism that creates confining strings inside domain walls. One of the two models presented in \cite{Auzzi:2008zd} is a supersymmetric theory that has important similarities with the one discussed in the present paper.

The supersymmetric theory considered in \cite{Auzzi:2008zd} is $\mathcal{N}=2$ gauge theory, 
with the gauge group ${\rm U}(2)= {\rm SU}(2) \times {\rm U}(1) /Z_2 $,
with no matter hypermultiplets. The following superpotential which breaks
the extended supersymmetry down to $\mathcal{N}=1$  is then added: 
\beq W= \alpha \,  \Tr \, \left(
\frac{\Phi^3}{3} - \xi \Phi \right). \label{superpotential} \eeq Classically, we have
three vacua, with $\phi$ equal to:
\beq  
\label{classicalvacua}
\left( \begin{array}{cc} \phantom{.}\sqrt{\xi} & 0\\[2mm]0  & \phantom{-}\sqrt{\xi} 
\end{array}\right) \ , \qquad 
\left( \begin{array}{cc} \phantom{.}\sqrt{\xi} & 0 \\[2mm]
0 & -\sqrt{\xi} \end{array}\right) \ , 
\qquad 
\left( \begin{array}{cc}- \sqrt{\xi} & 0\\[2mm]
0 &- \sqrt{\xi} \end{array}\right) \ .
   \label{threevacua}
\eeq   
The first and the last vacua preserve the non-Abelian SU(2) gauge symmetry. Strong coupling effects
\`{a} la Seiberg and Witten will then split each of them into two vacua (the monopole 
and dyon vacua). The vacuum in the middle preserves only the U$(1) \times {\rm U}(1)$ gauge symmetry,
and is not split. We, thus, expect in total  five vacua, for   generic values of $\xi$. The five vacua are depicted in Figure \ref{vacuaxibig}. The value of $u_2$ is  $\xi$ for all  five vacua. It is not modified by quantum corrections.  The Coulomb vacuum in the middle is not modified by quantum correction either.
\FIGURE[h]{
\includegraphics[width=18em]{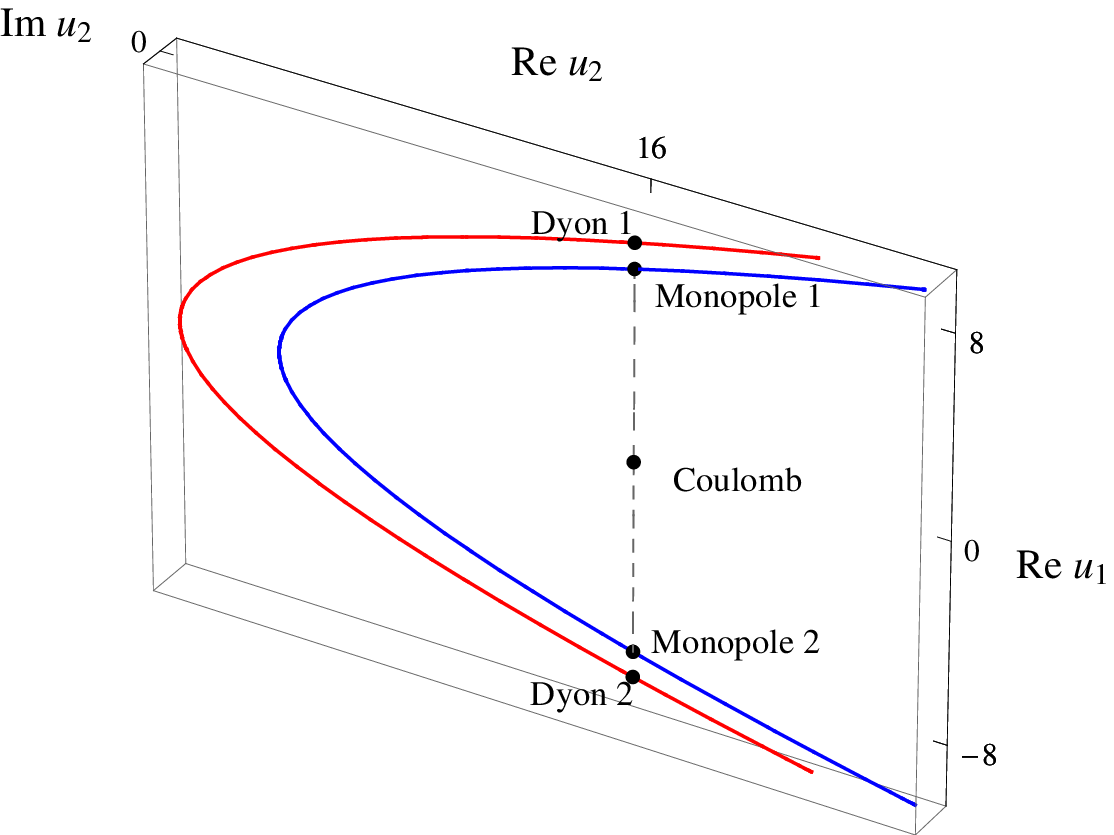}
\caption{ Five vacua of the model ($\Lambda=1$, $\xi=4$). The dashed line corresponds to the composite domain wall.}
\label{vacuaxibig}
}
In the limit $\sqrt{\xi} \gg \Lambda$ the Coulomb vacuum is such that the electric coupling is small. 
As $\xi$ decreases and becomes of order $\Lambda$,  the Coulomb vacuum enters a strong coupling regime. At the critical value  $\xi =  \Lambda^2 $  the Coulomb vacuum lies exactly in the monopole singularity and coalesces with  two monopole vacua.  Around this critical value, the  Coulomb vacuum is such that the magnetic coupling is small, so we can use the same set of low-energy effective variables
to describe both the Coulomb and the confining vacua.

The value of the superpotential in the two confining vacua,
monopole-1 and monopole-2,
is:
\beq
W = \pm \,
\frac{4}{3} \alpha  (\xi-\Lambda^2)^{3/2}\, . 
\label{vccv}
\eeq 
In the
Coulomb vacuum the superpotential vanishes. Hence,  the
BPS bound for the tension of the wall interpolating between monopole-1 and monopole-2 vacua,
if it existed,  would be
twice that of the BPS wall interpolating between the Coulomb
and  confining vacua. The latter walls will be referred to as elementary.
The former wall can be called composite.  The three values of the superpotential are allineated in the complex plane.

In this theoretical set-up, no BPS wall interpolating between  two
confining vacua, monopole-1 and monopole-2,    exists. In other words, a composite wall
built of two elementary walls at a finite distance from each other,
does not exist.
Supersymmetric solutions correspond to solutions of a dynamical system determined by  the first-order equations,
starting from monopole-1, following the profile $W$ and ending in
monopole-2.
A field configuration interpolating between monopole-1 and monopole-2 is
always time-dependent; it represents two elementary walls moving 
under the influence of a repulsive 
force between them (see Ref.~\cite{Portugues:2001ah}). 
This force falls off exponentially with the wall separation. Alternatively we can say that the composite wall exists and is BPS, but the distance between the elementary ones is infinite.

In order to avoid this problem and stabilize the composite domain wall, an extra term  is
introduced in the superpotential:
\beq 
W=  \alpha \left( {\rm Tr}  \left(
\frac{\Phi^3}{3} - \xi \Phi \right  ) + \frac{i \mu}{2} ({\rm
Tr} \,\Phi)^2 \right),
\label{modsp}
\eeq
where $\xi$ and $\mu$ are real mass parameters.
The values of the superpotential  in both confining vacua change, due to the double trace operator, of the same quantity. The Coulomb vacuum is instead unaffected.
The net result is that the three values of the superpotential are no more allineated in the complex plane.

The tension of the BPS domain wall is given by the absolute value of the difference of
the superpotentials at two vacua between which the given wall interpolates. 
For this reason, if the   composite BPS  walls
exist at $\mu \neq 0$, the composite wall will be stable.
Direct numerical investigation has revealed that the composite domain-wall is indeed BPS, and the distance between the two elementary walls is stabilized by the $\mu$ deformation.

We would like to understand the localization of the (massive) gauge field 
on the wall as a quasimodulus $\sigma$ localized on the wall world volume. 
The condition that typical energies in the low-energy theory
must be $  \ll  1/d$ cannot  be met then. 
In this formulation, it makes no sense to  speak of localization
and reduction to $2+1$ dimensions.
The wall at $\sigma =\pi$ correspond, in fact, to the infinite distance between the two elementary ones.

Thus, although the low-energy description in the Seiberg--Witten motivated model
at hand  is
not of the sine-Gordon type, 
the quasimodulus-based low-energy description is still
valid at $|\sigma |\ll \pi$: a mass term $m\sigma^2$ is generated.

The conclusion is that: {\it 1)} The present model is a realization of the string-inside-wall phenomenon; {\it 2)} This string can not be interpreted as a domain line in a $2+1$ effective action.

\vskip 0.50cm
\begin{center}
*  *  *
\end{center} 
\noindent
In the previous example, confinement inside domain walls is not a phenomenon that can be, in general, captured by a $2+1$ low energy effective action. This is always the case is the model under consideration shares a fundamental property with the example just described. The key properties are the following: $(i)$ the domain wall interpolates between two vacua where a common particle condenses and creates a vortex; $(ii)$ there is an intermediate vacuum, a \textit{true} one, where the particle is massive. In this case the confinement can not be understood from a $2+1$ effective action, of whatsoever kind.

The strings inside walls in $\N=1$ SYM fall exactly in this category.
Thus, the fact that Acharya-Vafa theory is in the topological massive phases and shows no sign of confining strings, this should not be viewed as a contradiction to our statement.

To understand better this last remark, let us refer to the example given in Section \ref{fieldtheory}, in particular the $2$-wall in the SU$(4)$ gauge theory. The composite operator  $\widetilde{E}_1^{\0} E_3^{\0}$ in the 0-vacuum  is exactly the same as the composite operator $\widetilde{E}_1^{\2} E_3^{\2}$ in the 2-vacuum. This is the scalar field responsible for the creation of the domain line living inside the domain wall. This is obtained by the winding of the  relative phase between the operator on the left side and on the right side (what we called $\sigma$ in Ref.~\cite{Auzzi:2008zd}), This phase {\it can not} be implemented on a $2+1$ effective action. This would be possible only if every configuration with constant $\sigma$ could be obtained as an adiabatic deformation of the lowest state ($\sigma=0$, the basic domain wall) while preserving the condition of validity of the effective action, that is energy$ \ll 1/d$ with $d$ the thickness of the wall. The configuration $\sigma =\pi$ is the one in which the condensates of $\widetilde{E}_1^{\0} E_3^{\0}$ and $\widetilde{E}_1^{\2} E_3^{\2}$
vanish in the middle of the composite wall. This can be obtained only when the two elementary walls are at infinite distance. Since $d \to \infty$ as $\sigma \to \pi$, the condition of validity of the effective action inevitably vanishes.  a low-energy effective action can only see the fluctuation of $\sigma$ around the zero and is inevitably blind to the topological structure which is essential to explain the formation of the string-wall bound state.

%% file: junction.tex
At the base of the present paper is the fact that a confining $1$-string can end on a $1$-wall in $\N=1$ SYM.   The heuristic interpretation is that the flux carried by the string is screened by any possible combination of condensates  at the two sides of the domain wall. If such a combination exists then we say that the string can terminate on the domain. Otherwise we have the string inside walls phenomenon described in the paper.

The purpose of this Appendix is to describe in more detail the string-wall junction and the bi-condensate effect that is responsible for the screening of the string flux.
To do this we work in the context of $\N=2$ broken by a small $\mu$ mass term. The two vacua are a
monopole $(0,1)$
and a
dyon $(2,-1)$,
that respectively condense and give confinement.

We describe the domain wall between the monopole ((0) vacuum) and the dyon ((1) vacuum) much on the same lines as has been done in \cite{Kaplunovsky:1998vt}. We neglect completely the boundary effects, and consider the condensate to drop to zero like a step function. There is an inner layer between both the condensates and the energy density is non-vanishing. We set equal to $1$ this non-zero energy density.  Stabilization of the wall is given introducing a scalar field $u$ which has value $-1$ on the $(0)$ side and $+1$ on the $(1)$ side. 
The tension as a function of the distance $d$ of the domain wall is $d+2/d$.
Stabilization is given by the balance of forces, given by the derivatives $-1+2/d^2=0$.   $1$ is the force, per unit of area, due to the energy density on the internal layer; it is constant and always inward (negative). The other term is a positive force (outward) equal to the derivative of the scalar field $u$.
The result is a stable domain wall with tension $T_{\W}=2\sqrt{2}$  and thickness $d=\sqrt2$.

In a generic setup, where the shape is not necessary that of a straight domain wall, the situation changes slightly, and essential  now is the use of the stabilization of forces argument.
The energy density always gives an inward force of magnitude one, both on the $(0)$ and $(1)$ sides. For the scalar field we have to solve the Laplacian equation with Dirichlet boundary conditions on the two sides:
\beq
\triangle u =0 \ , \qquad u|_{\0}=-1 \ , \qquad u|_{\1}=+1
\eeq
The force from the scalar field is given by the gradient $\vec{\nabla}u$ and directed outward..

A similar schematization can be used for the confining string. We consider the string in the monopole vacuum, and thus it will carry electric flux equal to $2 \pi$.
The tension of this vortex is the sum of the electric energy contribution,  $E^2/2$ times the area, plus the vacuum energy contribution, $1$ times the area. In total it is $2\pi / r^2 + \pi r^2$. The force per unit of area acting on the monopole surface is 
$(1/2\pi r)$ times the derivative $\partial_r$ of the previous expression: $-2/r^4 + 1=0$.
The balance of forces is thus $E^2/2 - 1=0$. Again, there is an inward force of magnitude $1$ due to the energy density. The other force is $E^2/2$ and is directed outward, and this is because the electric field is acting on the monopole current that creates it.
The result is a vortex with tension  $T_{\S}=2\sqrt2 \pi$ and radius  $r=\sqrt[4]{2}$.

We are now ready to combine the two objects, vortex and wall, together in the junction of Figure \ref{giunzione}. We use cylindrical coordinates $r,z,\theta$.
\FIGURE[h]{
\includegraphics[width=28em]{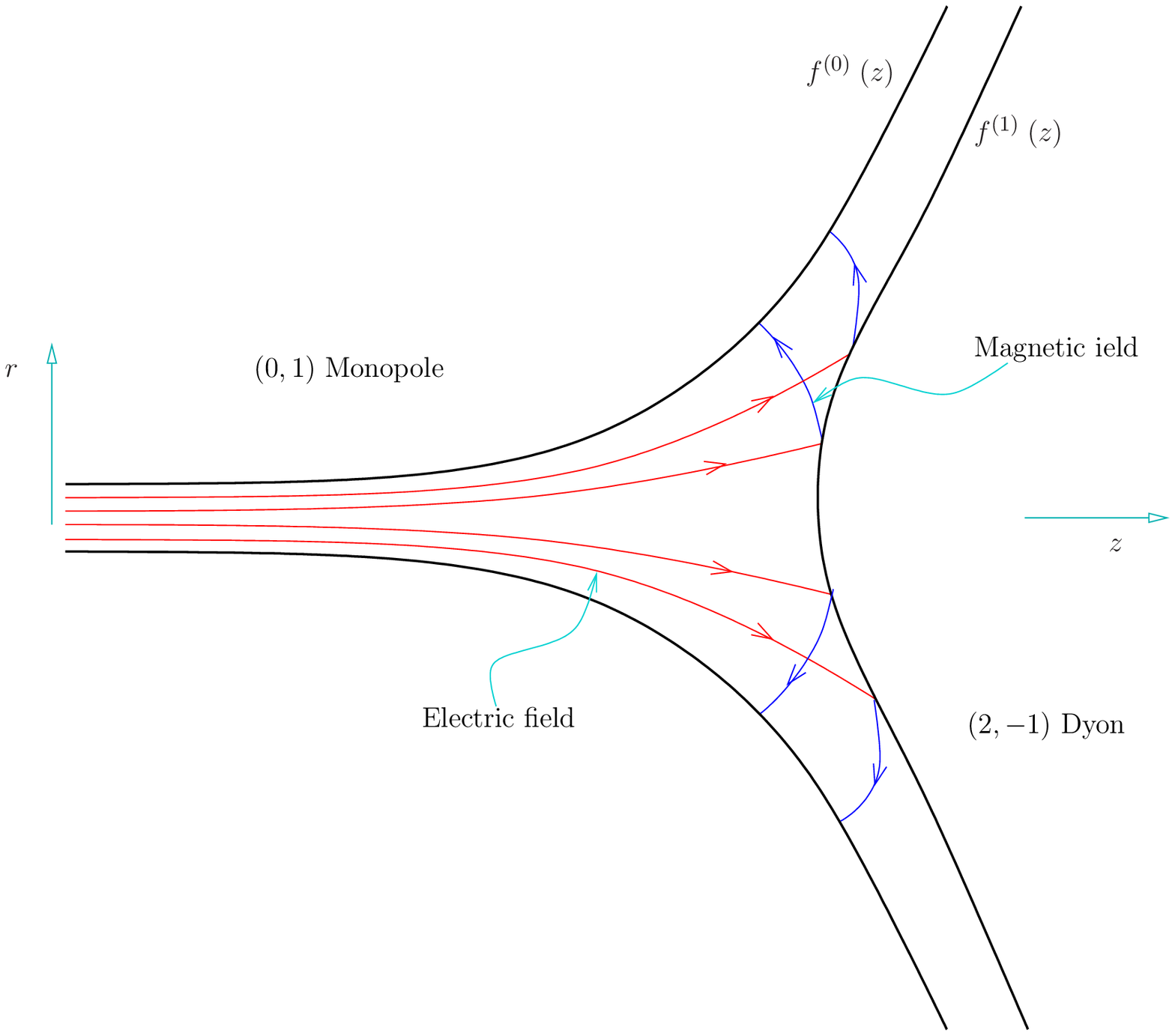}
\caption{ The string-wall junction in our basic approximation. }
\label{giunzione}
}
The vortex comes from the monopole side and is continuously  merged with the domain wall. In this crude approximation, the solution of the junction is given by the two profiles $r=f^{\0}(z)$ and $r=f^{\1}(r)$, respectively, the edge of the monopole condensate and the edge of the dyon condensate.  In the middle there are no condensates and the energy density acts as an inward force on every surface element of the monopole and dyon borders.

To obtain the other forces acting on the surfaces we have to solve the Laplace equation for three functions: $u,\varphi^{\e},\varphi^{\m}$.
The first is the scalar field already introduced before. 
The second and the third are, respectively, the electric and magnetic potentials: $\vec{E}=\vec{\nabla}\varphi^{\e}$, $\vec{B}=\vec{\nabla}\varphi^{\m}$.
Now the essential ingredients are the boundary conditions. On the monopole side the electric field must be tangential to the surface $f^{\0}$ (Neumann boundary condition) while the 
magnetic field must be perpendicular (Dirichlet boundary conditions):
\beq
\vec{n} \cdot \vec{\nabla} \varphi^{\e}|_{\0}=0 \ , \qquad \varphi^{\m}|_{\0}={\rm const}=0
\eeq
The electric field is generated by monopole current, like in the solitonic vortex. The force that the electric field exerts on the surface is {\it outward} with magnitude $E^2/2$. The magnetic field is generated by the monopole condensate itself, like in a solitonic Q-ball. 
To obtain the boundary condition on the dyon side we have to perform th correct duality transformation. The combination $2\varphi^{\e}-\varphi^{\m}$ is the one with Dirichlet boundary conditions, while $\varphi^{\e}+2\varphi^{\m}$ is the one  with Neumann boundary condition:
\beq
\vec{n} \cdot \vec{\nabla} (\varphi^{\e}+2\varphi^{\m})|_{\1}=0 \ , \qquad (2\varphi^{\e}-\varphi^{\m})|_{\1}={\rm const}=0
\eeq

Finally, we writhe the force balance condition on the two surfaces. On the monopole side we have that the force that the electric field exerts on the surface is {\it outward} with magnitude $E^2/2$. The force is again $B
^2/2$, but this time directed {\it inward}.
\beq
\left. \left( -1+\vec{\nabla}u+\frac{\vec{E}^2}{2} -\frac{\vec{B}^2}{2} \right) \right|_{\0} =0
\eeq
On the dyon surface we have a very similar equation but with the correct duality transformation:
\beq
\left. \left( -1+\vec{\nabla}u+\frac{(-\vec{E}+2\vec{B})^2}{2} -\frac{(2\vec{E}+\vec{B})^2}{2} \right) \right|_{\1}=0
\eeq

Up to now we have provided the equations that govern the string-wall junction. Solving them is not easy, it could in principle be done with a recursive numerical technic similar to the one used in \cite{Bolognesi:2006pp}.  We start with certain functions $f^{\0}$ and $f^{(\1)}$. Then we solve the Laplace equations with the given boundary conditions. We thus have the forces that act on the two surfaces. We then modify the functions in order to minimize the forces and reach zero at the end of the iteration. Although easy to say, it is a numerically challenging problem.

We have given the equations that determine the junction profiles $f^{\0}$ and $f^{\1}$, and we have given an implementable method to solve the equation numerically. 
The string-wall junction, from this solitonic point of view, is composed by a vortex, two Q-balls (the bi-condensate that screens the vortex flux) and a domain wall.
As a further step is necessary to understand how to connect the fields, even in the bulk where the monopole and the dyon condense. Non-locality is the main obstacle to this understanding. It is, in fact, clear that a global solution with a gauge potential $\varphi,\vec{A}$ is not possible.

%% file: main.bbl
\begin{thebibliography}{99}
{\small \itemsep -2pt }



\bibitem{SD}
  G.~R.~Dvali and M.~A.~Shifman,
  ``Domain walls in strongly coupled theories,''
  phys.\ Lett.\  B {\bf 396}, 64 (1997)
  [Erratum-ibid.\  B {\bf 407}, 452 (1997)]
  [arXiv:hep-th/9612128].
  %%CITATION = PHLTA,B396,64;%%

\bibitem{WittenMQCD}
  E.~Witten,
  ``Branes and the dynamics of {QCD},''
  Nucl.\ Phys.\ B {\bf 507}, 658 (1997)
  [arXiv:hep-th/9706109].


\bibitem{Mio}
  S.~Bolognesi,
  ``Domain walls and flux tubes,''
  Nucl.\ Phys.\  B {\bf 730} (2005) 127
  [arXiv:hep-th/0507273].
  %%CITATION = NUPHA,B730,127;%%

\bibitem{HSZ}
 A.~Hanany, M.~J.~Strassler and A.~Zaffaroni,
  ``Confinement and strings in M{QCD},''
  Nucl.\ Phys.\  B {\bf 513}, 87 (1998)
  [arXiv:hep-th/9707244].
  %%CITATION = NUPHA,B513,87;%%


\bibitem{Volo}
  A.~Volovich,
  ``Domain walls in M{QCD} and Monge-Ampere equation,''
  Phys.\ Rev.\  D {\bf 59} (1999) 065005
  [arXiv:hep-th/9801166].
  %%CITATION = PHRVA,D59,065005;%%

\bibitem{MN}
J.~M.~Maldacena and C.~Nunez,
  ``Towards the large N limit of pure N = 1 super Yang Mills,''
  Phys.\ Rev.\ Lett.\  {\bf 86}, 588 (2001)
  [arXiv:hep-th/0008001].
  %%CITATION = PRLTA,86,588;%%

\bibitem{KS}
 I.~R.~Klebanov and M.~J.~Strassler,
  ``Supergravity and a confining gauge theory: Duality cascades and
  chiSB-resolution of naked singularities,''
  JHEP {\bf 0008} (2000) 052
  [arXiv:hep-th/0007191].
  %%CITATION = JHEPA,0008,052;%%

\bibitem{wittenbaryons}
E.~Witten,
  ``Baryons and branes in anti de Sitter space,''
  JHEP {\bf 9807} (1998) 006
  [arXiv:hep-th/9805112].
  %%CITATION = JHEPA,9807,006;%%

\bibitem{acharyavafa}
 B.~S.~Acharya and C.~Vafa,
  ``On domain walls of N = 1 supersymmetric Yang-Mills in four dimensions,''
  arXiv:hep-th/0103011.
  %%CITATION = HEP-TH/0103011;%%

\bibitem{mishacounting}
 A.~Ritz, M.~Shifman and A.~Vainshtein,
  ``Counting domain walls in N = 1 super Yang-Mills,''
  Phys.\ Rev.\  D {\bf 66} (2002) 065015
  [arXiv:hep-th/0205083].
  %%CITATION = PHRVA,D66,065015;%%

\bibitem{topologicalmass}
S.~Deser, R.~Jackiw and S.~Templeton,
  ``Topologically massive gauge theories,''
  Annals Phys.\  {\bf 140}, 372 (1982)
  [Erratum-ibid.\  {\bf 185}, 406.1988\ APNYA,281,409 (1988\ APNYA,281,409-449.2000)].
  %%CITATION = APNYA,281,409;%%


\bibitem{polyakovconfinment}
  A.~M.~Polyakov,
  ``Quark Confinement And Topology Of Gauge Groups,''
  Nucl.\ Phys.\  B {\bf 120} (1977) 429.
  %%CITATION = NUPHA,B120,429;%%

\bibitem{Dvali:2007nm}
  G.~Dvali, H.~B.~Nielsen and N.~Tetradis,
  ``Localization of Gauge Fields and Monopole Tunnelling,''
  arXiv:0710.5051 [hep-th].
  %%CITATION = ARXIV:0710.5051;%%
  
  
\bibitem{Auzzi:2008zd}
  R.~Auzzi, S.~Bolognesi, M.~Shifman and A.~Yung,
  ``Confinement and Localization on Domain Walls,''
  arXiv:0807.1908 [hep-th].  
  
\bibitem{Witten:1999ds}
  E.~Witten,
  ``Supersymmetric index of three-dimensional gauge theory,'' arXiv:hep-th/9903005.
  %%CITATION = HEP-TH/9903005;%%

  


\bibitem{Campos:1998db}
  A.~Campos, K.~Holland and U.~J.~Wiese,
  ``Complete wetting in supersymmetric {QCD} or why {QCD} strings can end on
  domain walls,''
  Phys.\ Rev.\ Lett.\  {\bf 81}, 2420 (1998)
  [arXiv:hep-th/9805086].
  %%CITATION = PRLTA,81,2420;%%

 
  
  

\bibitem{Kaplunovsky:1998vt}
  V.~S.~Kaplunovsky, J.~Sonnenschein and S.~Yankielowicz,
  ``Domain walls in supersymmetric Yang-Mills theories,''
  Nucl.\ Phys.\  B {\bf 552}, 209 (1999)
  [arXiv:hep-th/9811195].
  %%CITATION = NUPHA,B552,209;%%


\bibitem{Dvali:1999pk}
  G.~R.~Dvali, G.~Gabadadze and Z.~Kakushadze,
  ``BPS domain walls in large N supersymmetric {QCD},''
  Nucl.\ Phys.\  B {\bf 562}, 158 (1999)
  [arXiv:hep-th/9901032].



\bibitem{Gabadadze:1999pp}
  G.~Gabadadze and M.~A.~Shifman,
  ``D-walls and junctions in supersymmetric gluodynamics in the large N  limit
  suggest the existence of heavy hadrons,''
  Phys.\ Rev.\  D {\bf 61}, 075014 (2000)
  [arXiv:hep-th/9910050].
  %%CITATION = PHRVA,D61,075014;%%




\bibitem{'thooftexoticphases}
 G.~'t Hooft,
  ``On The Phase Transition Towards Permanent Quark Confinement,''
  Nucl.\ Phys.\  B {\bf 138} (1978) 1.
  %%CITATION = NUPHA,B138,1;%%




\bibitem{donagiwitten}
 R.~Donagi and E.~Witten,
  ``Supersymmetric Yang-Mills Theory And Integrable Systems,''
  Nucl.\ Phys.\  B {\bf 460} (1996) 299
  [arXiv:hep-th/9510101].
  %%CITATION = NUPHA,B460,299;%%



\bibitem{mishayung}
  M.~Shifman and A.~Yung,
  ``Domain walls and flux tubes in N = 2 SQCD: D-brane prototypes,''
  Phys.\ Rev.\  D {\bf 67} (2003) 125007
  [arXiv:hep-th/0212293].
  %%CITATION = PHRVA,D67,125007;%%

\bibitem{Isozumi:2004vg}
  Y.~Isozumi, M.~Nitta, K.~Ohashi and N.~Sakai,
  ``All exact solutions of a 1/4 Bogomol'nyi-Prasad-Sommerfield equation,''
  Phys.\ Rev.\  D {\bf 71}, 065018 (2005)
  [arXiv:hep-th/0405129].

  
\bibitem{tongsakay}
 N.~Sakai and D.~Tong,
  ``Monopoles, vortices, domain walls and D-branes: The rules of
interaction,''
  JHEP {\bf 0503} (2005) 019
  [arXiv:hep-th/0501207].
  %%CITATION = JHEPA,0503,019;%%

  
  
  \bibitem{bojium}
R.~Auzzi, M.~Shifman and A.~Yung,
  ``Studying boojums in N = 2 theory with walls and vortices,''
  Phys.\ Rev.\  D {\bf 72} (2005) 025002
  [arXiv:hep-th/0504148].
  %%CITATION = PHRVA,D72,025002;%%


\bibitem{robertostrings}
 R.~Auzzi, M.~Shifman and A.~Yung,
  ``Domain lines as fractional strings,''
  Phys.\ Rev.\  D {\bf 74} (2006) 045007
  [arXiv:hep-th/0606060].
  %%CITATION = PHRVA,D74,045007;%%


\bibitem{AHW}
  I.~Affleck, J.~A.~Harvey and E.~Witten,
  ``Instantons And (Super)Symmetry Breaking In (2+1)-Dimensions,''
  Nucl.\ Phys.\  B {\bf 206}, 413 (1982).
  %%CITATION = NUPHA,B206,413;%%

\bibitem{DS}
  M.~R.~Douglas and S.~H.~Shenker,
  ``Dynamics of SU(N) supersymmetric gauge theory,''
  Nucl.\ Phys.\  B {\bf 447} (1995) 271
  [arXiv:hep-th/9503163].



  \bibitem{Bolognesi:2004da}
S.~Bolognesi,
  ``The holomorphic tension of vortices,''
  JHEP {\bf 0501}, 044 (2005)
  [arXiv:hep-th/0411075].
  


  \bibitem{Armoni:2005sp}
  A.~Armoni and T.~J.~Hollowood,
  ``Sitting on the domain walls of N = 1 super Yang-Mills,''
  JHEP {\bf 0507} (2005) 043
  [arXiv:hep-th/0505213].
  %%CITATION = JHEPA,0507,043;%%

  \bibitem{Armoni:2006ee}
  A.~Armoni and T.~J.~Hollowood,
  ``Interactions of domain walls of SUSY Yang-Mills as D-branes,''
  JHEP {\bf 0602} (2006) 072
  [arXiv:hep-th/0601150].
  %%CITATION = JHEPA,0602,072;%%


\bibitem{Portugues:2001ah}
R.~Portugues and P.~K.~Townsend, ``Intersoliton forces in the Wess-Zumino model,'' Phys.\ Lett.\ B{\bf 530},
227 {2002} [arXiv:hep-th/0112077].


\bibitem{Bolognesi:2006pp}
S.~Bolognesi and S.~B.~Gudnason,
``Soliton junctions in the large magnetic flux limit,''
Nucl.\ Phys..\ B {\bf 754}, 293 (2006)
[arXiv:hep-th/0606065].


\end{thebibliography}
